\documentclass[onecolumn,usenatbib,amsmath,amssymb,amsbsy]{mn2e}

\usepackage{amsbsy}
\usepackage{amssymb}
\usepackage{epsfig,graphicx}
\usepackage{natbib}
\title[Two-fluid models of superfluid neutron star cores]{Two-fluid models of superfluid neutron star cores}
\author[N. Chamel]{N. Chamel\thanks{E-mail:
nchamel@ulb.ac.be}\\
Institut d'Astronomie et d'Astrophysique, Universit\'e Libre de Bruxelles, CP226, Boulevard du Triomphe, B-1050 Brussels, Belgium\\}

\pagerange{\pageref{firstpage}--\pageref{lastpage}} \pubyear{2008}

\begin{document}

\label{firstpage}

\maketitle

\begin{abstract}
Both relativistic and non-relativistic two-fluid models of neutron star cores are constructed, using the constrained variational formalism developed by Brandon Carter and co-workers. We consider a mixture of superfluid neutrons and superconducting protons at zero temperature, taking into account mutual entrainment effects. Leptons, which affect the interior composition of the neutron star and contribute to the pressure, are also included. We provide the analytic expression of the Lagrangian density of the system, the so-called master function, from which the dynamical equations can be obtained. All the microscopic parameters of the models are calculated consistently using the non-relativistic nuclear energy density functional theory. For comparison, we have also considered relativistic mean field models. The correspondence between relativistic and non-relativistic hydrodynamical models is discussed in the framework of the recently developed 4D covariant formalism of Newtonian multi-fluid hydrodynamics. We have shown that entrainment effects can be interpreted in terms of dynamical effective masses that are larger in the relativistic case than in the Newtonian case. With the nuclear models considered in this work, we have found that the neutron relativistic effective mass is even greater than the bare neutron mass in the liquid core of neutron stars.

\end{abstract}


\section{Introduction}
\label{sect.intro}

The recent discovery of quasi periodic oscillations (QPOs) in the X-ray flux of giant flares from 
soft-gamma repeaters (SGR) may well be the first direct observational evidence of neutron star oscillations.
QPOs have been detected during the 2004 December 27 giant flare from SGR 1806-20~\citep{israel-05, strohmayer-06, watts-06}, 
during the 1998 August 27 giant flare from
SGR 1900+14~\citep{strohmayerw-05} and during the 1979 March 5 event in SGR 0526-66~\citep{barat-83}. 
Those QPOs are usually interpreted as global seismic vibrations triggered by magnetic crust quakes (for
 a recent review of these so-called magnetars, see for instance~\citealt{woods-06} and references therein). 
If this interpretation is confirmed, the analysis of those QPOs can potentially reveal the interior 
composition of neutron stars, thus putting constraints on the theory of dense matter~\citep{samuelsson-07}. 
Apart from QPOs in SGR, the rapid development of the gravitational wave astronomy opens very exciting perspectives of 
directly observing neutron star oscillations in a near future~\citep{andersson-05}. In particular, 
accreting neutron stars in Low Mass X-Ray Binaries are expected to be detectable by the Advanced LIGO detector\footnote{http://www.ligo.caltech.edu/advLIGO/}
which is planned to be operational in a few years~(see for instance the recent analysis of \citealt{watts-08}). 
However the interpretation of these observations requires not only a detailed theoretical understanding of the dynamics 
of neutron stars, but also a consistent description of the different layers. 

A neutron star is mainly composed of 
three distinct regions: an outer crust, an inner crust characterized by the presence of a neutron ocean and a 
liquid core which might be solid in the deepest regions~\citep*{haensel-06}. 
Microscopic calculations of dense nuclear matter suggest that the matter inside neutron stars is superfluid~\citep{dean-03}. 
This theoretical prediction is strongly supported by the observations of  pulsar glitches~\citep{baym-69,anderson-75}. 
Other indications in favour of superfluidity, while less convincing, are provided by observations of neutron star thermal X-ray
emission~\citep{yakovlev-04}. One of the remarkable consequences of superfluidity is the possibility of having several 
dynamically distinct components. The electrically charged particles inside neutron stars are locked together by the interior 
magnetic field and co-rotate on very long time scales of the order of the age of the star~\citep{easson-79}. The charged particles 
are rotating at the observed angular velocity of the star due to the coupling with the radiating magnetosphere and thus follow 
the long-term spinning-down of the star caused by the electromagnetic radiation. In contrast the neutrons being electrically 
uncharged and superfluid can rotate at a different rate. This naturally leads to considering the interior of a neutron star as a two-fluid mixture. 
As a result of the strong interactions between neutrons and protons, the two fluids are not completely independent but are coupled 
via mutual entrainment effects. These non-dissipative effects are known to affect significantly the frequencies of 
superfluid oscillation modes for which the superfluid neutrons and the charged particles are counter moving~\citep{andersson-01}. 
Two-fluid models of superfluid neutron star cores including entrainment effects, have been proposed by \citet{comer-03}. \citet*{cchII-05, cchI-06} 
(see also~\citealt{chamelcarter-06}) have shown how to describe in a unified way, both the liquid core and the inner crust within 
this two-fluid picture. The description of the outer crust requires a different treatment~\citep*{carter-06}. Eventually the 
different layers have to be matched with appropriate boundary conditions at the interfaces~\citep{andersson-02, lin-07}. 

In this work, we have constructed relativistic hydrodynamical models of cold superfluid neutron star cores, 
calculating \emph{all} the necessary microscopic coefficients with the \emph{same} underlying microscopic model. 
The present work differs from that of \citet{comer-03} by improving the microphysics description of dense nuclear matter 
and by making the link between non-relativistic and relativistic models. 
In the first Section, we briefly review the convective variational formalism of multi-fluid systems. This approach is 
employed to construct a two-fluid model of neutron star core in the Newtonian framework in Section~\ref{sect.non-rel.hydro}. 
It is then shown in Section~\ref{sect.rel.hydro} how to generalize this model to relativistic fluids. In Section~\ref{sect.effmass}, 
we discuss the effects of entrainment in terms of dynamical effective masses. 
The consequences of superfluidity 
at the hydrodynamical scale are discussed in Section~\ref{sect.super}. Section~\ref{sect.composition} is devoted to the conditions 
of ``chemical'' equilibrium and to the composition of neutron star cores. In Section~\ref{sect.micro}, the microscopic parameters 
of the two-fluid model are evaluated, using Skyrme effective nucleon-nucleon interactions. As an example, numerical results 
are shown in Section~\ref{sect.examples} for three particular Skyrme forces: the popular SLy4 force and the parameter sets LNS and NRAPR which were 
entirely constructed from realistic quantum many body calculations. In Section~\ref{sect.rmf}, results are compared to those obtained by the relativistic mean field theory applied by \citet{comer-03}. We have constructed new relativistic mean field models that yield a much better 
agreement with nuclear data than those considered by \citet{comer-03}. 

\section{Variational formalism of multi-fluid hydrodynamics}
\label{sect.variation}

In the following, we will use Greek letters for the space-time indices $\mu, \nu, \dots$ and Latin letters $i,j, \dots$ for space indices. 
We introduce capital Latin letters $_{\rm X}, _{\rm Y}$ for distinguishing the various constituents and among them we will adopt the symbols 
$\ell$ for leptons and $q$ for nucleons (when several indices of the same species will be needed, we will add primes on the label as for instance $q$, $q^\prime$, $q^{\prime\prime}$, etc.). We will apply the Einstein summation convention (i.e. repeated indices are summed) for space-time 
indices but not for the constituent labels. 

Let us consider an arbitrary number of fluids that are interacting with each other. We follow the variational formalism developed by~\citet{carter-89}, which has been recently reviewed by~\citet{gourgoulhon-06} and~\citet{anderssoncomer-07}. In this approach, the basic fluid variables are the particle 4-currents $n_{_{\rm X}}^{\, \mu}$ of each fluid. 
The equation governing the dynamical evolution of each fluid is obtained from an action principle by considering 
variations of the fluid particle trajectories. Given a so-called master function $\Lambda$, which is the Lagrangian density of the system, 
and a set of 4-force densities $f^{_{\rm X}}_{\,\nu}$ acting on each fluid, the hydrodynamic equations take the very simple form
\begin{equation}
\label{eq.Euler}
n_{_{\rm X}}^{\, \mu}\varpi^{_{\rm X}}_{\!\mu\nu} + \pi^{_{\rm X}}_{\, \nu}\nabla_\mu n_{_{\rm X}}^{\, \mu}  = f^{_{\rm X}}_{\,\nu} \, ,
\end{equation}
where the vorticity 2-form $\varpi^{_{\rm X}}_{\!\mu\nu}$ is defined as the exterior 
derivative of the 4-momentum covector
\begin{equation}
\label{eq.4momentum}
\pi^{_{\rm X}}_{\, \mu} = \frac{\partial \Lambda}{\partial n_{_{\rm X}}^{\, \mu}} \, ,
\end{equation}
namely
\begin{equation}
\label{eq.vorticity}
\varpi^{_{\rm X}}_{\!\mu\nu}= 2\nabla_{\![\mu}\pi^{_{\rm X}}_{\,\nu]} = \nabla_{\!\mu}\pi^{_{\rm X}}_{\,\nu} - \nabla_{\!\nu}\pi^{_{\rm X}}_{\,\mu} \, ,
\end{equation}

It is understood that the partial derivative in Eq.~(\ref{eq.4momentum}) is taken with all other 4-currents being kept constant. 
For a strict application of the variational principle, the forces should separately vanish $f^{_{\rm X}}_{\,\nu}=0$, which entails by contracting 
Eq.~(\ref{eq.Euler}) with the corresponding 4-current $n_{_{\rm X}}^{\, \nu}$, that $\nabla_\mu n_{_{\rm X}}^{\, \mu}=0$. In this case, 
Eq.~(\ref{eq.Euler}) therefore reduce to 
\begin{equation}
\label{eq.Euler2}
n_{_{\rm X}}^{\, \mu}\varpi^{_{\rm X}}_{\!\mu\nu} = 0 \, .
\end{equation}
Note that despite the vanishing of the individual forces $f^{_{\rm X}}_{\,\nu}$, the fluids are not independent of each other in general and the set of Eqs~(\ref{eq.Euler2}) are therefore not simply Euler equations. The couplings between the various fluids are hindered in the momenta $\pi^{_{\rm X}}_{\, \mu}$. 

The stress-energy tensor of the fluids can be expressed as 
\begin{equation} 
\label{eq.stress.tensor}
T^\mu_{\ \nu}=\Psi\,\delta^\mu_\nu + \sum_{^{\rm X}} n_{_{\rm X}}^{\, \mu} \pi^{_{\rm X}}_{\, \nu} \, ,
\end{equation}
where $\Psi$ can be interpreted as a generalized pressure and is defined by
\begin{equation} 
\label{eq.general_pressure}
\Psi=\Lambda-\sum_{^{\rm X}} n_{_{\rm X}}^{\, \mu} \pi^{_{\rm X}}_{\, \mu} \, .
\end{equation}
Let us emphasize that so far we have made no assumption with respect to the space-time geometry so that the 
above covariant expressions, based only on the exterior calculus, are valid both in (special and general) relativity and in the Newtonian limit. 
In the following sections we will show how to construct the Lagrangian density $\Lambda$ in each case.

\section{Non-relativistic two-fluid models of neutron star core}
\label{sect.non-rel.hydro}

We consider a uniform mixture with four constituents: neutrons, protons, electrons and possibly muons. 
Such a composition is expected to be found in the interior of low-mass neutron stars and in the outer core of massive neutron stars at 
densities above the crust-core transition density $\rho_{\rm cc}\sim \rho_0/2$ and below $\lesssim 2-3 \rho_0$, where 
$\rho_0 \simeq 2.8\times 10^{14}$ g.cm$^{-3}$ is the saturation density of infinite symmetric nuclear matter. Given the current uncertainties 
on the composition of neutron star core (for a recent review see for instance~\citealt*{haensel-06}), it is 
sometimes assumed for simplicity that this composition remains the same at higher densities. 

At densities below $\sim 3 \rho_0$, the nucleons are essentially non-relativistic. 
For instance, the sound velocity for the realistic model $A18+\delta v+{\rm UIX}^*$ of~\citet{akmal-98}, 
becomes comparable to the speed of light at densities around $\sim 5 \rho_0$. For this model, such densities are 
reached in very massive neutron stars with a mass larger than $2 M_\odot$. However the most precisely measured 
neutron star masses (in neutron star binaries) lie below $1.5 M_\odot$~\citep{lattimer-07}. Besides as shown 
by~\citet{glendenning-00} (chapter 3, Section 4), the effects of General relativity are negligible at the microscopic scale.
Moreover at the macrocoscopic scale, the fluid velocities
are small compared to the speed of light. Indeed the velocity at the equator of the most rapidly spinning neutron stars is only about 
$\sim 20\%$ of the speed of light (taking $1$ ms for the period and $10$ km for the radius). 
For the purpose of matching the microscopic nuclear model to the macroscopic hydrodynamical model, it will therefore be convenient to start 
with a local analysis considering non-relativistic fluids in the Newtonian framework. Such non-relativistic models can be also 
very useful by themselves for studying qualitatively the dynamics of superfluid mixtures in neutron stars~\citep{andersson-01}. In order to facilitate the correspondence between relativistic and non-relativistic models, we will use the fully 4D covariant formalism developed by~\citet{CCI-04,CCII-05,CCIII-05}. We will then show how to construct fully relativistic fluid models in Section~\ref{sect.rel.hydro} (as required 
for a General Relativistic description of the star).
For simplicity we only consider the possible presence of the magnetic field by supposing that leptons and protons 
are co-moving, as discussed in Section~\ref{sect.intro}. We therefore consider only two independent fluids: the neutron superfluid and 
the fluid of charged particles (protons, electrons and possibly muons). This two-fluid model includes the limit of 
non-superfluid neutron star cores since in this case, all the particles are essentially co-moving and can thus be treated as a single
fluid~\citep{baym-69}. Including the magnetic field is in principle straightforward. It has been recently shown that under some circumstances 
the magnetic field can even be variationally taken into account in the purely Newtonian context despite the non-Galilean invariance 
of Maxwell's equations~\citep{carter-06}. However taking into account the magnetic field as a dynamical field, implies a better understanding of the 
lepton dynamics as well as the proton superconductivity which is beyond the scope of the present model. 

Let us now briefly review the 4-dimensional geometric structure of the Newtonian space-time (see~\citealt{CCI-04} for a detailed discussion). Newtonian theory postulates the existence of a universal time $t$, leading to a foliation of the space-time into 3-dimensional hypersurfaces. Each of these spatial sections are flat and are endowed with the 3-dimensional Euclidean metric, giving rise to the symmetric contravariant tensors $\eta^{\mu\nu}$ and $\eta_{\mu\nu}$ by pushforward and by pull-back respectively. These tensors are not metric tensors since they are degenerate 
\begin{equation}
\eta^{\mu\nu} t_\nu = 0 \, , \hskip1cm \eta_{\mu\nu} e^\nu = 0
\end{equation}
where $t_\nu = \partial_\nu t$ and the ``ether'' flow vector $e^\mu$, normalized by the condition
\begin{equation}
\label{eq.norm_ether}
e^\mu t_\mu = 1 \, ,
\end{equation}
characterizes a particular Aristotelian frame corresponding to the usual kind of 3+1 space-time decomposition.
The particle 4-currents introduced in Section~\ref{sect.variation} are given by $n_{_{\rm X}}^{\, \mu}=n_{_{\rm X}} u_{_{\rm X}}^{\, \mu}$, where $n_{_{\rm X}}$ are the corresponding particle number densities, and the 4-velocities $u_{_{\rm X}}^{\, \mu}$ are defined by
\begin{equation}
\label{eq.u}
u_{_{\rm X}}^{\, \mu}=v_{_{\rm X}}^{\, \mu} + e^\mu\, , \hskip1cm t_\mu v_{_{\rm X}}^{\, \mu}=0\, .
\end{equation}
with $v_{_{\rm X}}^{\, \mu}$ being the corresponding push forward of the usual 3-velocities in the given Aristotelian frame. 
It is easily seen from Eq.~(\ref{eq.norm_ether}) that the 4-velocities are normalized as 
\begin{equation}
\label{eq.norm_u}
t_\mu u_{_{\rm X}}^{\, \mu}=1\, .
\end{equation}

We will neglect the small mass difference between neutrons and protons and write simply $m$ for the nucleon mass, which we take equal to the atomic mass unit. For consistency we thus take the electron mass $m^e=0$. We also neglect the muon mass except at the microscopic scale 
for calculating the internal energy density.
In the limit of small currents, the Lagrangian density $\Lambda$ can be decomposed into a dynamical part $\Lambda_{\rm dyn}$, which depends on the particle currents $n_{_{\rm X}}^{\, \mu}$, and a static part $\Lambda_{\rm ins}$ which depends only on the particle densities given by
\begin{equation}
\label{eq.densities}
n_{_{\rm X}}=n_{_{\rm X}}^{\, \mu} t_\mu \, . 
\end{equation}
The dynamical Lagrangian density (neglecting the contribution of the leptons) can be written as a quadratic form of the nucleon currents
\begin{equation}
\label{eq.lambda.dyn}
 \Lambda_{\rm dyn} = \frac{1}{2} \sum_{q, q^\prime} \eta_{\mu\nu}\, {\cal K}^{q q^\prime} n_{q}^{\, \mu} n_{q^\prime}^{\, \nu} \, .
\end{equation}
The coefficients of the symmetric mobility matrix ${\cal K}^{q q^\prime}$ ($q,q^\prime=n,p$ for neutrons, protons respectively) are 
functions of the nucleon densities and will be evaluated in Section~\ref{sect.micro}. Due to the Galilean invariance, the matrix elements are related to each other by 
\begin{equation}\label{eq.gal.inv}
\sum_{q^\prime} n_{q^\prime} {\cal K}^{q q^\prime} = m \, .
\end{equation}
Taking the partial derivative of the above equation with respect to the nucleon density leads to the following 
identity
\begin{equation}\label{eq.gal.inv2}
\sum_{q^\prime,q^{\prime\prime}} n_{q^\prime} \frac{\partial{\cal K}^{q^\prime q^{\prime\prime}}}{\partial n_q} n_{q^{\prime\prime}} = -m \, .
\end{equation}
Since the left-hand side of Eq.(\ref{eq.gal.inv2}) has to be negative for any neutron and proton densities, we obtain the following inequalities
\begin{equation}
\frac{\partial {\cal K}^{qq}}{\partial n_{q^\prime}} <0 \, , \hskip1cm \left(\frac{\partial {\cal K}^{np}}{\partial n_q}\right)^2 <\frac{\partial {\cal K}^{nn}}{\partial n_q}\frac{\partial {\cal K}^{pp}}{\partial n_q} \, .
\end{equation}
The non-diagonal coefficient ${\cal K}^{np}={\cal K}^{pn}$ accounts for the non-dissipative entrainment effects and arises from the strong interactions between nucleons. If the nucleons could be treated as ideal Fermi gases, this coefficient would simply vanish ${\cal K}^{np}=0$. 

The static contribution $\Lambda_{\rm ins}$ is related to the static internal (non gravitational) energy density $U_{\rm ins}$ by
\begin{equation}
\label{eq.lambda.ins}
\Lambda_{\rm ins} = - U_{\rm ins} - U_{\rm pot} \, ,
\end{equation}
where $\rho$ is the total baryon mass density
\begin{equation}
\rho = m n_{\rm b} \, 
\end{equation}
and $U_{\rm pot}=\rho \phi$ is the gravitational potential energy density,
$\phi$ being the gravitational scalar potential. 
As discussed previously nucleons are not relativistic neither at the 
microscopic scale nor at the macroscopic scale. Nevertheless, in order to prepare the generalization to relativistic fluids in the next section, 
the nucleon rest mass energy density is included in $U_{\rm ins}$. It can be shown that in the Newtonian framework this extra term 
has no effect on the variational principle (thereby on the dynamical equations of the fluids) provided the total mass current is conserved~\citep{carter-06}. Let us point out that the total internal (non gravitational) energy density $U_{\rm int}$ is not simply given 
by $U_{\rm ins}$, but contains an additional entrainment term defined by $U_{\rm ent}= U_{\rm dyn} - U_{\rm kin}$, where 
$U_{\rm dyn}=\Lambda_{\rm dyn}$ and 
\begin{equation}
U_{\rm kin}= \frac{1}{2} \sum_{q}  \eta_{\mu\nu}\, \frac{m}{n_{q}} n_{q}^{\, \mu} n_{q}^{\, \nu} \, ,
\end{equation}
is the total kinetic energy density. We thus have $U_{\rm int}=U_{\rm ins}+U_{\rm ent}$.

In summary, the non-relativistic Lagrangian density of the fluids is therefore given by
\begin{equation}
\label{eq.lambda}
\Lambda=\Lambda_{\rm dyn}+\Lambda_{\rm ins}\, ,
\end{equation}
where $\Lambda_{\rm dyn}$ and $\Lambda_{\rm ins}$ are given by Eqs.~(\ref{eq.lambda.dyn}) and (\ref{eq.lambda.ins}) respectively. 
The dynamical equations of the gravitational field can be obtained from the variational principle by simply 
adding the contribution
\begin{equation}
\label{eq.lambda.grf}
\Lambda_{\rm grf}=-\frac{1}{8\pi G} \eta^{\mu\nu}(\nabla_{\mu}\phi)(\nabla_{\nu}\phi)
\end{equation}
$G$ being the gravitational constant. Considering variations of the \emph{total} Lagrangian density $\Lambda_{\rm tot}=\Lambda+\Lambda_{\rm grf}$ 
with respect to $\phi$ leads to the usual Poisson's equation
\begin{equation}
\eta^{\mu\nu}\nabla_\mu\nabla_\nu \phi = 4\pi G\rho \, .
\end{equation}

Given the above Lagrangian density~(\ref{eq.lambda}), the momentum of each constituent is obtained from Eq.~(\ref{eq.4momentum}). 
The nucleon momentum is given by
\begin{equation}
\label{eq.nucleon_momentum}
\pi^{q}_{\, \mu} = \sum_{q^\prime} \eta_{\mu\nu}\, {\cal K}^{qq^\prime} n_{q^\prime}^{\, \nu} + t_\mu\left( \frac{1}{2}\sum_{q^\prime, q^{\prime\prime}} \eta_{\rho\nu} \frac{\partial {\cal K}^{q^\prime q^{\prime\prime}}}{\partial n_{q} } n_{q^\prime}^{\, \rho} n_{q^{\prime\prime}}^{\, \nu} - \mu_q - m \phi\right) \, ,
\end{equation}
where $\mu_{_{\rm X}}$ is the chemical potential defined by
\begin{equation}
\label{eq.muX}
\mu_{_{\rm X}} = \frac{\partial U_{\rm ins}}{\partial n_{_{\rm X}}}\, .
\end{equation}
Since the Lagrangian density depends only on the lepton densities $n_\ell=n_\ell^\mu t_\mu$ ($\ell=e,\mu$ for electrons, muons respectively),
the momenta of the leptons are time-like  and are given by
\begin{equation}
\label{eq.lepton_momentum}
\pi^\ell_{\, \mu} = - t_\mu \mu_\ell \, .
\end{equation}
If the particles are all co-moving with the 4-velocity $u^\mu$, 
the nucleon 4-momenta take the familiar expression
\begin{equation}
\label{eq.piq.comoving}
\pi^{q}_{\, \mu} = m \eta_{\mu\nu} u^\nu - t_\mu \left( \frac{1}{2} m v^2 +\mu_q + m \phi\right)
\end{equation}
where $v^2=\eta_{\mu\nu} u^\nu u^\mu$,
using the identities~(\ref{eq.gal.inv}) and (\ref{eq.gal.inv2}). 

The usual 3-momentum covector, denoted by $\Pi^{_{\rm X}}_{\, \mu}$, is defined by the Aristotelian spatial components of the 4-momentum covector $\pi^{_{\rm X}}_{\, \mu}$, 
\begin{equation}
\label{eq.3pi}
\Pi^{_{\rm X}}_{\, \mu}=\eta^\mu_\nu \pi^{_{\rm X}}_{\, \nu} \, ,
\end{equation}
where $\eta^\mu_\nu$ is the space projection tensor defined by 
\begin{equation}
 \eta^\mu_\rho = \delta^\mu_\nu - e^\mu t_\nu \, ,
\end{equation}
using the Kronecker unit tensor $\delta^\mu_\nu$. It follows immediately from Eq.~(\ref{eq.norm_ether}) that $e^\mu \Pi^{_{\rm X}}_{\, \mu}=0$. 
It is readily seen that the lepton 3-momentum vanishes $\Pi^{\ell}_{\, \mu}=0$, while the nucleon 3-momentum is determined solely by the mobility matrix ${\cal K}^{qq^\prime}$
\begin{equation}
\Pi^{q}_{\, \nu}= \sum_{q^\prime} \eta_{\nu\mu}\, {\cal K}^{qq^\prime} n_{q^\prime}^{\, \mu}\, .
\end{equation}

From the nucleon and lepton momenta Eqs.~(\ref{eq.nucleon_momentum}) and (\ref{eq.lepton_momentum}) respectively, we can obtain the 
generalized pressure $\Psi$ of the fluids according to Eq.~(\ref{eq.general_pressure}). 
The kinetic part of the Lagrangian density $\Lambda_{\rm kin} = U_{\rm kin}$
does not contribute to the pressure $\Psi$ which can thus be written as 
\begin{equation}
\Psi=\Lambda_{\rm int} - \sum_{^{\rm X}} n_{_{\rm X}}^{\, \mu}\,\frac{\partial \Lambda_{\rm int}}{\partial n_{_{\rm X}}^{\, \mu}} \, ,
\end{equation}
where $\Lambda_{\rm int}= \Lambda-\Lambda_{\rm kin}$. Due to entrainment effects, this internal Lagrangian density $\Lambda_{\rm int}$ 
of the fluids is not simply equal to the opposite of the \emph{total} internal energy density (including gravitational contribution) 
$U_{\rm int}+U_{\rm pot}$
 but is given by 
\begin{equation}
\Lambda_{\rm int} = -U_{\rm int} -U_{\rm pot} +2 U_{\rm ent}\, .
\end{equation}
The gravitational potential energy density does not contribute to the pressure. 
As a result, the generalized pressure $\Psi$ is the sum of a static term $\Psi_{\rm ins}$ given by 
\begin{equation}
\label{eq.psi.static}
\Psi_{\rm ins}=\sum_{^{\rm X}} n_{_{\rm X}}\frac{\partial U_{\rm ins}}{\partial n_{_{\rm X}}} - U_{\rm ins} \, ,
\end{equation}
and a dynamical term given by
\begin{equation}
\Psi_{\rm ent}=-\sum_{^{\rm X}} n_{_{\rm X}}^{\, \mu}\,\frac{\partial U_{\rm ent}}{\partial n_{_{\rm X}}^{\, \mu}} + U_{\rm ent} \, .
\end{equation}
In the single fluid case, the static pressure $\Psi_{\rm ins}$ reduces to the ordinary pressure usually denoted by $P$. Note that 
in multi-fluid systems, if the static internal energy density is of the form 
\begin{equation}
U_{\rm ins}=\sum_{^{\rm X}} U^{_{\rm X}}\{ n_{_{\rm X}} \}
\end{equation}
the static pressure can then be written as the sum of the partial pressures $P_{_{\rm X}}$ 
\begin{equation}
P=\sum_{^{\rm X}} P_{_{\rm X}} 
\end{equation}
with
\begin{equation}
P_{_{\rm X}} = \sum_{^{\rm X}} n_{_{\rm X}}\frac{\partial U^{_{\rm X}}}{\partial n_{_{\rm X}}} - U^{_{\rm X}} \, .
\end{equation}
In the general case however such a decomposition is not possible. 
The additional pressure term $\Psi_{\rm ent}$ vanishes whenever either the two fluids are co-moving or the fluids are non-interacting so that the non-diagonal coefficients of the mobility matrix vanish (no entrainment). 

Going back to the two-fluid model of neutron star cores, the static and entrainment contributions to the general pressure are given explicitly by
\begin{equation}
\label{eq.ord.P}
\Psi_{\rm ins}=n_n\frac{\partial U_{\rm ins}}{\partial n_n}+n_p\frac{\partial U_{\rm ins}}{\partial n_p}+n_e\frac{\partial U_{\rm ins}}{\partial n_e}+n_{\mu}\frac{\partial U_{\rm ins}}{\partial n_{\mu}} - U_{\rm ins} \, 
\end{equation}
and
\begin{equation}
\Psi_{\rm ent}=\frac{1}{2} \eta_{\mu\nu}(v_p^\mu-v_n^\mu)(v_p^\nu-v_n^\nu)\left[ n_n^2\frac{\partial m_\star^n}{\partial n_n} + n_p^2\frac{\partial m_\star^p}{\partial n_p} + \frac{1}{2} n_n (m_\star^n-m) + \frac{1}{2} n_p (m_\star^p-m)\right] \, ,
\end{equation}
respectively.
Given the general pressure $\Psi=\Psi_{\rm ent}+\Psi_{\rm ins}$ and the 4-momenta $\pi^{_{\rm X}}_{\, \mu}$ of the various constituents, the stress-energy tensor of the fluids can easily be obtained from Eq.~(\ref{eq.stress.tensor}).

\section{From non-relativistic to relativistic two-fluid models}
\label{sect.rel.hydro}

In the previous section, we have taken into account of the effects of the gravitational field on the fluids by including the term $-\rho \phi$ in the static internal Lagrangian density $\Lambda_{\rm ins}$. Alternatively as was first shown by Elie Cartan, the effects of gravitation can be taken into account in the structure of the Newtonian space-time itself, thus facilitating the comparison with General Relativity (see for instance chapter 12 of ~\citealt{mtw-73}). This can be achieved by replacing the tensor $\eta_{\mu\nu}$ in Eq.~(\ref{eq.lambda.dyn}) for the Newton-Cartan space-time metric $\gamma_{\mu\nu}$ defined by
\begin{equation}
\gamma_{\mu\nu} = \eta_{\mu\nu} - 2 \phi t_\mu t_\nu \, .
\end{equation}
It is readily verified, remembering Eq.~(\ref{eq.gal.inv}), that this is indeed completely equivalent to adding 
the term $-\rho \phi$ to the Lagrangian density $\Lambda$. However using this approach, it becomes clear how to make the correspondence with 
General Relativity (thereby Special Relativity as well). For weak gravitational fields, the Riemannian metric $g_{\mu\nu}$ of the relativistic space-time can be locally approximated by
\begin{equation}
g_{\mu\nu} \simeq \eta_{\mu\nu} - (c^2 + 2\phi)t_\mu t_\nu =\gamma_{\mu\nu} - c^2 t_\mu t_\nu \, .
\end{equation}
This suggests to define the ``dynamical'' contribution to the relativistic Lagrangian density as
\begin{equation}
\label{eq.lambda.dyn.rel}
\widetilde{\Lambda}_{\rm dyn} = \frac{1}{2} \sum_{q,q^\prime}  {\cal K}^{qq^\prime}\, \left(g_{\mu\nu}n_{q}^{\, \mu} n_{q^\prime}^{\, \nu} + c^2 n_{q} n_{q^\prime}\right) \, ,
\end{equation}
where $n_{q}^{\, \mu}=n_{q}u_q^{\, \mu}$. The particle densities $n_{_{\rm X}}$ are now defined by 
\begin{equation}
\label{eq.rel.densities}
n_{_{\rm X}} = \sqrt{-g_{\mu\nu} n_{_{\rm X}}^{\, \mu} n_{_{\rm X}}^{\, \nu}}/c\, ,
\end{equation}
with the 4-velocities normalized as 
\begin{equation}
\label{eq.norm_u.rel}
g_{\mu\nu} u_{_{\rm X}}^{\, \mu} u_{_{\rm X}}^{\, \nu} = -c^2 \, .
\end{equation}
Note that Eq.~(\ref{eq.rel.densities}) with the normalisation~(\ref{eq.norm_u.rel}) are consistent with Eqs.~(\ref{eq.u}) and 
Eq.~(\ref{eq.norm_u}) in the non-relativistic limit. Likewise with the definition~(\ref{eq.lambda.dyn.rel}), the 
expression~(\ref{eq.lambda.dyn}) is indeed recovered in the Newtonian limit.

The relativistic expression of the corresponding ``static'' part is readily obtained by simply substituting the particle densities of the constituents in the Newtonian ether frame for the densities~(\ref{eq.rel.densities}) in the rest frame of the corresponding particles. Using the identity~(\ref{eq.gal.inv}) and adding the internal energy density, the total relativistic Lagrangian density of the fluids can be expressed as 
\begin{equation}
\label{eq.lambda.rel}
\widetilde{\Lambda}=\frac{1}{2} \sum_{q,q^\prime}  {\cal K}^{qq^\prime}\, g_{\mu\nu}n_{q}^{\, \mu} n_{q^\prime}^{\, \nu}  +\frac{1}{2} n_{\rm b} m c^2 - U_{\rm ins}  \, ,
\end{equation}
where $n_{\rm b}=n_n+n_p$ is the baryon density. If the fluids are described in General Relativity, the Lagrangian~(\ref{eq.lambda.rel}) has to be complemented with the Einstein-Hilbert contribution
\begin{equation}
\widetilde{\Lambda}_{\rm grf}=\frac{c^4}{16\pi G} R\, ,
\end{equation}
where $R$ is the Ricci scalar associated with the metric $g_{\mu\nu}$. Variations of the action with respect to the metric (which involve not only variations of the total Lagrangian density but also variations of the space-time measure) leads to Einstein's equations (see for instance~\citealt{anderssoncomer-07}). 

The relativistic momenta of the nucleons can be expressed in a form similar to Eq.~(\ref{eq.3pi}) as 
\begin{equation}
\label{eq.piq.rel}
 \pi^{q}_{\, \mu} = \sum_{q^\prime} g_{\mu\nu}\widetilde{\cal K}^{qq^\prime}  n_{q^\prime}^{\, \nu} \, .
\end{equation}
The non-diagonal components of the symmetric relativistic mobility matrix $\widetilde{\cal K}^{q q^\prime}$, are equal to those of 
the non-relativistic matrix ${\cal K}^{qq^\prime}$, while the diagonal elements are given by 
\begin{equation}
\label{eq.kappa.rel}
\widetilde{\cal K}^{q q} = {\cal K}^{qq} - \frac{m}{n_q} + \frac{\mu_q}{c^2 n_{q}} -\frac{1}{c^2 n_{q}}\frac{\partial {\cal K}^{np}}{\partial n_{q} } \left( c^2 n_n n_p +g_{\rho\sigma}n_{n}^{\, \rho} n_{p}^{\, \sigma}\right)  \, .
\end{equation}
The relativistic momenta of the leptons take a very simple form
\begin{equation}
\label{eq.pil.rel}
\pi^{\ell}_{\, \mu} = \frac{\mu_\ell}{c^2} u_p^\mu\, g_{\mu\nu} \, .
\end{equation}
If the constituents are all co-moving with the 4-velocity $u^\mu$, the 4-momenta reduce to 
\begin{equation}
\label{eq.pi.rel.comoving}
\pi^{_{\rm X}}_{\, \mu} =\frac{\mu_{_{\rm X}}}{c^2} u^\mu\, g_{\mu\nu} \, .
\end{equation}

With the momenta specified, we can obtain the generalized pressure $\Psi$ from Eq.~(\ref{eq.general_pressure}). As for the non-relativistic case, $\Psi$ can be decomposed into 
an ordinary ``static'' part given by Eq.~(\ref{eq.psi.static}) and an extra contribution $\Psi_{\rm ent}$ due to entrainment 
which can be expressed as 
\begin{equation}
\Psi_{\rm ent}=-\sum_{^{\rm X}} n_{_{\rm X}}^{\, \mu}\,\frac{\partial \widetilde{U}_{\rm ent}}{\partial n_{_{\rm X}}^{\, \mu}} + \widetilde{U}_{\rm ent} \,  ,
\end{equation}
where the entrainment energy density is now defined by 
\begin{equation}
\widetilde{U}_{\rm ent} =\frac{1}{2} \sum_{q,q^\prime}  {\cal K}^{qq^\prime}\,  g_{\mu\nu}n_{q}^{\, \mu} n_{q^\prime}^{\, \nu} \, .
\end{equation}

Before concluding this section, let us remark that the relativistic Lagrangian density can be written in the very concise form 
\begin{equation}
\label{eq.lambda.rel2}
\widetilde{\Lambda} = \lambda_0 + \lambda_1 (x^2 - n p) \, ,
\end{equation}
where the coefficients $\lambda_0$ and $\lambda_1$ are given by 
\begin{equation}
\lambda_0=-U_{\rm ins} \, , \hskip0.5cm \lambda_1=-c^2 {\cal K}^{np} \, ,
\end{equation}
and adopting the following notations 
\begin{equation}
x^2 c^2 = -g_{\mu\nu}n_n^{\, \mu} n_p^{\, \nu}\, ,
\end{equation}
\begin{equation}
n^2 c^2 =  -g_{\mu\nu}n_n^{\, \mu} n_n^{\, \nu}\, ,
\end{equation}
\begin{equation}
p^2 c^2 =  -g_{\mu\nu}n_p^{\, \mu} n_p^{\, \nu} \, .
\end{equation}
Equation~(\ref{eq.lambda.rel2}) is consistent with the expansion of the Lagrangian density in powers of $(x^2 - n p)$, suggested 
by~\citet*{andersson-02}.
It can be clearly seen on Eq.~(\ref{eq.lambda.rel2}), that in the absence of entrainment (i.e. $\lambda_1=0$)  or in the case of co-moving fluids, the Lagrangian density reduces to the opposite of the internal energy density.

\section{Dynamical effective masses}
\label{sect.effmass}

\subsection{Non-relativistic case}

If the nucleons were not interacting with each other, the mobility matrix introduced in Section~\ref{sect.non-rel.hydro} 
would be diagonal and we would simply have ${\cal K}^{nn}=m/n_n$ and ${\cal K}^{pp}=m/n_p$. Of course we know that nucleons
are strongly interacting. This means that the matrix ${\cal K}^{qq^\prime}$ does not have such a simple structure. 
It is convenient to define neutron and proton dynamical effective masses by
\begin{equation}
\label{eq:effmass}
m_\star^n\equiv n_n\, {\cal K}^{nn}\, , \hskip 0.5cm m_\star^p\equiv n_p\, {\cal K}^{pp}\, 
\end{equation}
respectively. The deviations of $m_\star^q$ from the bare baryon mass $m$ therefore arise entirely from the nucleon-nucleon 
interactions. Let us point out that these effective masses depend on the nucleon densities and therefore vary with depth 
inside the neutron star. 
As a result of Eq.~(\ref{eq.gal.inv}), the non-diagonal coefficients of the mobility matrix can be expressed as 
\begin{equation}
\label{eq.mobility.matrix}
{\cal K}^{np}={\cal K}^{pn}=\frac{m-m_\star^n}{n_p}=\frac{m-m_\star^p}{n_n}\, .
\end{equation}
It can be shown with stability arguments~\citep{chamelhaensel-06} that the effective masses are bounded from below 
$m_\star^q/m>n_q/n_{\rm b}$, where $n_{\rm b}=n_n+n_p$ or equivalently 
\begin{equation}
\label{eq.stability}
{\cal K}^{np}={\cal K}^{pn}<\frac{m}{n_{\rm b}}\, .
\end{equation}
Since in neutron star core the effective masses are typically \emph{smaller} than the bare nucleon mass (therefore ${\cal K}^{np}>0$), 
this inequality provides an upper bound for the largest possible strength of entrainment effects between the two fluids. 
The physical meaning of the dynamical effective masses defined by Eq~(\ref{eq:effmass}), becomes clear when writing the 
expressions of the nucleon 3-momentum covectors~(\ref{eq.3pi})
\begin{equation}
\Pi^{n}_{\, \nu}=  m_\star^n v_{n\, \nu} + (m-m_\star^n) v_{p\, \nu}\, ,
\end{equation}
\begin{equation}
\Pi^{p}_{\, \nu}=  m_\star^p v_{p\, \nu} + (m-m_\star^p) v_{n\, \nu}\, .
\end{equation}
where $v_{_{\rm X}\, \nu}\equiv\eta_{\nu\mu}v_{_{\rm X}}^{\, \mu}$. This shows that the 3-momentum and the 3-velocity of 
a given nucleon species are not aligned whenever the dynamical effective masses differ from the bare nucleon mass, 
or equivalently whenever the non-diagonal coefficients of the mobility matrix do not vanish. 

\subsection{Relativistic case}

By analogy with the definition~(\ref{eq:effmass}), let us introduce relativistic nucleon dynamical  
effective masses by
\begin{equation}
\label{eq:releffmass}
\widetilde{m}_\star^q = n_q \widetilde{\cal K}^{q q}\, ,
\end{equation}
where $\widetilde{\cal K}^{q q^\prime}$ is the relativistic generalisation of the non-relativistic mobility 
matrix ${\cal K}^{qq^\prime}$. 
Using Eq.~(\ref{eq.kappa.rel}) together with (\ref{eq:effmass}), we find 
\begin{equation}
\label{eq:releffmass2}
\frac{\widetilde{m}_\star^q}{m} = \frac{m_\star^q}{m} + \frac{\mu_q}{m c^2} - 1 - \frac{1}{m c^2}\frac{\partial {\cal K}^{np}}{\partial n_{q} } \left( c^2 n_n n_p +g_{\rho\sigma}n_{n}^{\, \rho} n_{p}^{\, \sigma}\right)  \, .
\end{equation}
with $\mu_q$ the chemical potential defined by Eq.~(\ref{eq.muX}).
It is easily checked that $\widetilde{m}_\star^q \rightarrow m_\star^q$ in the Newtonian limit. 
With these definitions, the neutron and proton 4-momenta can be explicitly written as 
\begin{equation}
\pi^{n}_{\, \mu} =  \left[ \widetilde{m}_\star^n u_n^\nu + (m-m_\star^n) u_p^\nu \right] g_{\mu\nu}\, ,
\end{equation}
\begin{equation}
\pi^{p}_{\, \mu} =  \left[ \widetilde{m}_\star^p u_p^\nu + (m-m_\star^p) u_n^\nu \right] g_{\mu\nu}\, .
\end{equation}
Note that the entrainment contributions involve the \emph{non-relativistic} effective masses~(\ref{eq:effmass}). 
Eq.~(\ref{eq:releffmass2}) is the generalization to interacting multi-fluid systems of the effective mass introduced by \citet{carter-89}
in the perfect fluid case. Indeed in the absence of entrainment, ${\cal K}^{np}=0$ so that $m_\star^q=m$ while the relativistic effective masses
are given by 
\begin{equation}
\frac{\widetilde{m}_\star^q}{m} =\frac{\mu_q}{m c^2}   \, .
\end{equation}
This equation is identical to Eq.(1.66) in the lectures notes of \citet{carter-89}. 
The physical origin of the difference between $\widetilde{m}_\star^q$ and $m_\star^q$ is that in relativity 
\emph{all} forms of energy contribute to the mass. A remarkable consequence 
is that even massless particles can have a non-vanishing relativistic effective mass. 
This is for instance the case of leptons for which we have assumed $m^\ell=0$. However Eq.~(\ref{eq.pil.rel}) show that 
the dynamical effective lepton mass is not zero but is given by 
\begin{equation}
\widetilde{m}_\star^\ell =\frac{\mu_\ell}{c^2}   \, .
\end{equation}
Note that even if leptons were not interacting (which we will actually suppose in Section~\ref{sect.micro} in order 
to evaluate the master function $\Lambda$), they would still have an non-zero effective mass due to the Pauli exclusion principle
which prevents all the particles from occupying the lowest energy state with zero momentum (the chemical potential $\mu_\ell$ 
is then given by the Fermi energy of the lepton species $\ell$). 

For unbound nuclear systems like the liquid core of neutron stars, we have
$\mu_q> mc^2$. Besides if we assume that the strength of entrainment effects decreases with increasing density, i.e. 
\begin{equation}
\frac{\partial {\cal K}^{np}}{\partial n_{q} } < 0\, ,
\end{equation}
(this is actually the case for the models considered in this work, see Eq.~(\ref{eq.kappanp})),
and using the Cauchy-Schwartz inequality
\begin{equation}
-g_{\mu\nu} n_{n}^{\, \mu} n_{p}^{\, \nu} \leq \sqrt{-g_{\mu\nu} n_{n}^{\, \mu} n_{n}^{\, \nu}}\sqrt{-g_{\mu\nu} n_{p}^{\, \mu} n_{p}^{\, \nu}}
% c^2 n_n n_p +g_{\rho\sigma}n_{n}^{\, \rho} n_{p}^{\, \sigma} >0\, ,
\end{equation}
it is easily shown that the relativistic effective masses are always larger than the non-relativistic ones. 
With the notations introduced at the end of Section~\ref{sect.rel.hydro}, the relativistic dynamical effective masses 
can be expressed solely in terms of the parameters $\lambda_0$ and $\lambda_1$ of the relativistic Lagrangian density as
\begin{equation}
\label{eq:releffmass3}
\widetilde{m}_\star^q = (n_{\rm b}-n_q) \frac{\lambda_1}{c^2}-\frac{1}{c^2}\frac{\partial \lambda_0}{\partial n_q}+\frac{1}{c^2} \frac{\partial \lambda_1}{\partial n_q} (n p -x^2) \, .
\end{equation}
Note that the first order expansion of the relativistic Lagrangian density, Eq.~(\ref{eq.lambda.rel2}) in powers of $(x^2- n p)$ leads to 
dynamical effective masses with a first order contribution proportional to $(x^2- n p)$. This velocity-dependent term vanishes 
in the Newtonian limit so that the non-relativistic effective masses are independent of the velocities to first order. 
The reason is that in relativity the particle number densities involve all the components of the 
corresponding 4-currents according to Eq.~(\ref{eq.rel.densities}) while in the Newtonian case, the densities only depend 
on the time component of the 4-current through Eq.~(\ref{eq.densities}).

\section{Superfluidity}
\label{sect.super}

In the previous sections, we have accounted for the superfluidity in neutron star core 
by assuming that two independent fluid motions could co-exist. However strictly speaking this assumption only requires perfect fluidity, 
i.e. the absence of viscosity and dissipative drag effects which damp the development of relative motions between the 
constituents. The distinguishing feature of a superfluid compared to a perfect fluid is the fact that it can be described 
by a macroscopic quantum wave function. This entails that in a superfluid the momentum circulation is quantized according to 
the Bohr-Sommerfeld quantization rule
\begin{equation}
\label{eq.super}
\oint \pi_{\, \mu} {\rm d}x^\mu = N  \pi \hbar \, ,
\end{equation}
where $\hbar$ is the Dirac-Planck constant and $N$ is an integer, which simply follows from the 
requirement that the length of any closed path must be an integral multiple of the de 
Broglie wavelength of the condensate formed of bound neutron pairs. This condition implies the existence of neutron quantized vortex lines in 
neutron stars~\citep{ginzburg-64}. Assuming that the neutron vortices are arranged on a 
regular triangular array, the inter vortex spacing is given by
\begin{equation}
\label{eq.intervortex}
d_\upsilon=\sqrt{\frac{h}{2 \sqrt{3} m \Omega_n}}\simeq 3.4\times 10^{-3} \sqrt{\frac{10^2\, {\rm s}^{-1}}{\Omega}}\, {\rm cm} \, ,
\end{equation}
where $\Omega_n$ is the angular velocity of the neutron superfluid, which is approximately equal to the observed angular 
velocity $\Omega$ of the star. 

In regions devoid of vortices, the neutron momentum circulation is equal to zero which implies that 
the neutron momentum can be written as 
\begin{equation}
\label{eq.super.pi}
\pi^{n}_{\, \mu} = \frac{\hbar}{2}\nabla_\mu \varphi^n \, ,
\end{equation}
where the factor of $1/2$ accounts for the fermionic nature of the neutrons and $\varphi^n$ is the scalar phase of the 
condensate. Consequently, the corresponding vorticity 2-form \emph{locally} vanishes
\begin{equation}
\label{eq.super.local}
\varpi^{n}_{\!\mu\nu}=0 \, .
\end{equation}
However at length scales much larger than $d_\upsilon$ for which we are interested here, a fluid element is 
threaded by many vortex lines. Consequently the vorticity 2-forms do not have to vanish at this scale. 
Nevertheless, the superfluidity condition requires the existence of an average 4-velocity vector $u_\upsilon^\mu$ of 
the vortex lines such that the Lie derivative of the vorticity 2-form along $u_\upsilon^\mu$ vanishes
\begin{equation}
\vec u_\upsilon \pounds \varpi^{n}_{\!\mu\nu}=0 \, .
\end{equation}
The above condition is satisfied if 
\begin{equation}
\label{eq.super.macro}
 u_\upsilon^\mu\varpi^{n}_{\!\mu\nu}=0 \, .
\end{equation}
Vortices are co-moving with the superfluid, unless forces act on them. Indeed it can be seen that 
Eq.~(\ref{eq.super.macro}) with $u_\upsilon^\mu=u_n^\mu$ is consistent with Euler Eq.~(\ref{eq.Euler2})
obtained for the case $f^n_\nu=0$. 
It should be stressed that the above conditions~(\ref{eq.super.local}) and (\ref{eq.super.macro}) 
apply for either relativistic or non-relativistic superfluid. The presence of vortices can be explicitly 
included in the variational principle as shown by \citet{carter-99} and will not be further discussed here.

\section{Composition of neutron star core}
\label{sect.composition}

The composition of the neutron star core is determined by the rates of transfusion processes 
which convert particles of different species into each other. While the baryon number is always conserved
\begin{equation}
\label{eq.baryon.cons}
\nabla_\mu n_b^\mu = 0 \, ,
\end{equation}
where $n_{\rm b}^\mu = n_n^\mu+n_p^\mu$, neutrons may be transformed into protons and \textit{vice et versa} 
via electroweak processes. The fastest process is the direct Urca process
\begin{equation}
\label{eq.durca.reactions}
n \rightarrow p^{+} + \ell + \bar \nu_\ell \, , \hskip 0.5cm p^{+}+\ell \rightarrow n + \nu_\ell \, ,
\end{equation}
where $\ell$ is electron or muon. 
When the beta equilibrium is reached, the two reactions occur at the same rate. 
In degenerate dense matter, this process is allowed for sufficiently large proton fractions owing to the 
requirement that both momentum and energy has to be conserved~\citep{lattimer-91}. When these reactions are forbidden, the slower modified Urca 
process prevails
\begin{equation}
\label{eq.murca.reactions}
n + N \rightarrow p^{+} + \ell + \bar \nu_\ell \, , \hskip 0.5cm p^{+}+N+ \ell \rightarrow n + \nu_\ell \, ,
\end{equation}
involving an additional spectator nucleon $N$ (neutron or proton). 
The relaxation time of these beta processes for $npe$ matter, neglecting nucleon superfluidity, is approximately given by
$\tau^{(D)} \sim 20 T_9^{-4}$ s and $\tau^{(M)} \sim T_9^{-6}$ months for the direct and modified Urca processes
respectively, where $T_9$ is the temperature in units of $10^9$ K~\citep{yakovlev-01}.
Electrons and muons are transformed into each other via 
the lepton modified Urca processes
\begin{equation}
\label{eq.lepton.murca.reactions}
e^{-} + X \rightarrow \mu^{-} + X + \bar \nu_\mu + \nu_e \, , \hskip 0.5cm \mu^{-} + X \rightarrow e^{-}+X + \bar\nu_e +\nu_\mu\, ,
\end{equation}
where $X$ is either a nucleon or a lepton (the direct process is kinematically forbidden). 
The relaxation time associated with electromagnetic processes, of the order of $10^{-22}$ s~\citep{easson-79b}, is much smaller than the 
characteristic time-scales of the neutron star phenomena considered here, so that the matter can be treated as
 electrically neutral. This condition reads
\begin{equation}
\label{eq.neutrality}
n_p=n_e+n_\mu \, .
\end{equation}

In the newly-born proto-neutron stars, the temperatures are of the order of $\sim 10^{11}$ K or higher so that the equilibrium is 
reached in a few microseconds for the modified Urca process or ten times less for the direct Urca. As the star cools down to temperatures 
$\sim 10^{9}$ K after $10^3-10^4$ years, the relaxation times rise dramatically to about 20 seconds for the direct Urca and several months 
for the modified Urca. Besides when the temperature falls below the critical threshold for the onset of superfluidity, 
the relaxation times increase exponentially~\citep{villain-05}. As a consequence, for the short time-scales $\sim 1-100$ milliseconds relevant for oscillations of mature neutron stars (like the recently observed QPOs in SGR), the composition of the star 
remains essentially frozen and the constituents can therefore be assumed to be separately conserved
\begin{equation}
\label{eq.current.cons}
\nabla_\mu n_p^\mu  = 0 \, , \hskip 0.5cm \nabla_\mu n_e^\mu  = 0 \, ,
\end{equation}
which entails by Eqs.~(\ref{eq.baryon.cons}) and (\ref{eq.neutrality}), that the other currents are 
also conserved $\nabla_\mu n_n^\mu  = 0$ and $\nabla_\mu n_\mu^\mu  = 0$ (remembering that the leptons are co-moving 
with the protons). Let us remark that the lepton number is not conserved 
unlike the baryon number, because the neutron star matter is transparent to neutrinos (except for the first few seconds after 
its birth into a hot proto-neutron star).

The initial equilibrium composition of the neutron star core is obtained from the condition of electro neutrality~(\ref{eq.neutrality})
and the conditions that the chemical affinities corresponding to the above processes should vanish~\citep{CCIII-05}. 
The chemical affinity ${\cal A}^{_\Xi}$ of a given reaction $\Xi$ 
is defined by~\citep{CCIII-05} 
\begin{equation}\label{eq.def.Axi}
{\cal A}^{_\Xi}= -\sum_{^{\rm X}} N^{_\Xi}_{_{\rm X}} {\cal E}^{_{\rm X}} \, ,
\end{equation} 
where $N^{_\Xi}_{_{\rm X}}$ and ${\cal E}^{_{\rm X}}$ are the relevant particle creation numbers and 
the energies per particle, respectively. 
As pointed out by~\citet{CCIII-05}, the problem arises of determining the reference frame with 
respect to which the energies ${\cal E}^{_{\rm X}}$ have to be measured when some of the constituents 
(here neutrons and charged particles) are moving with different velocities. Since the relative velocity 
between the two fluids is expected to be small compared to the fluid velocities, we assume for simplicity in this section  that 
the particles are all co-moving with 4-velocity $u^\mu$. It is then natural to define the energy per particle as 
${\cal E}^{_{\rm X}}=-u^\mu \pi^{_{\rm X}}_{\, \mu}$. In the Newtonian case, using Eq.~(\ref{eq.piq.comoving}) 
we thus have 
\begin{equation}
\label{eq.enpart}
{\cal E}^{_{\rm X}}=\mu_{_{\rm X}}+m^{_{\rm X}} \phi - \frac{1}{2} m^{_{\rm X}} v^2 \, ,
\end{equation}
where the chemical potential of a particle species $_{\rm X}$ is defined by Eq.~(\ref{eq.muX}). 
Since the mass is conserved in any chemical reaction $\Xi$ involving \emph{non-relativistic} particles, the corresponding chemical affinity~(\ref{eq.def.Axi})
 reduces to
\begin{equation}
\label{eq.Axi}
{\cal A}^{_\Xi}=-\sum_{^{\rm X}} N^{_\Xi}_{_{\rm X}}\mu_{_{\rm X}}  \, .
\end{equation} 
In the relativistic case, the energy per particle obtained from Eq.~(\ref{eq.pi.rel.comoving}) is given by 
\begin{equation}
\label{eq.enpart.rel}
{\cal E}^{_{\rm X}}=\mu_{_{\rm X}} \, ,
\end{equation}
so that Eq.~(\ref{eq.Axi}) is valid for both relativistic and non-relativistic particles. 

We assume that neutrinos have escaped from the star so that we set ${\cal E}^{\nu_\ell}=0$. 
Both the direct and modified Urca processes, respectively (\ref{eq.durca.reactions}) and 
(\ref{eq.murca.reactions}), have the same affinity given by 
\begin{equation}
{\cal A}^{\rm Urca}={\cal E}^n - {\cal E}^p-{\cal E}^e \, .
\end{equation} 
The electron-muon transfusion reactions~(\ref{eq.lepton.murca.reactions}) are characterised by the affinity
\begin{equation}
{\cal A}^{\mu e}={\cal E}^e -{\cal E}^{\mu} \, .
\end{equation} 
The composition of the core at a given baryon density $n_{\rm b}$, can then be determined by solving 
the equations ${\cal A}^{\rm Urca}=0={\cal A}^{\mu e}$ under the constraint~(\ref{eq.neutrality}). 
Using~(\ref{eq.Axi}), this leads to
\begin{equation}
\label{eq.beta.equi}
\mu_n=\mu_e+\mu_p \, ,
\end{equation}
\begin{equation}
\label{eq.beta.equi2}
\mu_e=\mu_\mu \, .
\end{equation}

If the matter is in equilibrium, the static pressure 
depends only on the baryon density $n_{\rm b}$ and can be written in the concise form (valid in both the relativistic case
and the Newtonian limit)
\begin{equation}
\label{eq.Peq}
\Psi_{\rm ins}=n_{\rm b} \mu_{\rm n}  -U_{\rm ins}
\end{equation}
where $\mu_{\rm n}$ is the neutron chemical potential evaluated at the \emph{equilibrium} neutron and proton densities, 
associated with the baryon density $n_{\rm b}$. Let us emphasize that Eq.~(\ref{eq.Peq}) is only valid for neutron star 
matter in \emph{equilibrium}. In the general case, the static pressure is given by Eq.~(\ref{eq.psi.static}). 

\section{Evaluation of the microscopic parameters}
\label{sect.micro}

The internal static energy density can be decomposed into several contributions
\begin{equation}
U_{\rm ins}\left\{n_n,n_p,n_e,n_{\mu}\right\} = U_{\rm N}\left\{n_n,n_p\right\} + U_{\rm Coul}\left\{n_p,n_e,n_{\mu}\right\}  + U_{\rm L}\left\{n_e\right\} + U_{\rm L}\left\{n_{\mu}\right\}  \, ,
\end{equation}
where $U_{\rm N}$ is the nucleon part, $U_{\rm Coul}$ is the Coulomb part and $U_{\rm L}$ the lepton (kinetic) part. 
The Coulomb energy arises from lepton-lepton, lepton-proton and proton-proton interactions. It is of purely 
quantum origin since the classical contribution coming from Poisson's equation vanishes as a result of 
electro neutrality. We neglect the lepton-lepton interactions (but not the lepton-proton interactions) and we approximate 
the proton Coulomb energy by the Hartree-Fock exchange energy of non-relativistic point-like charged particles
\begin{equation}
U_{\rm Coul}\left\{n_p\right\}=-\frac{3}{4}e^2\left(\frac{3}{\pi}\right)^{1/3} n_p^{4/3} \, .
\end{equation}
Unlike nucleons, leptons are relativistic at the microscopic scale (note however that this does not imply 
that their \emph{collective} motion is relativistic at the \emph{macroscopic} scale of the fluid description).
Their kinetic energy is thus given by that of an ideal relativistic Fermi gas ($\ell=e,\mu$)
\begin{equation}
\label{eq:lepton.energy}
U_{\rm L}\left\{n_\ell\right\} = \frac{\hbar c}{8 \pi^2\lambda_\ell^4}\left[ x_\ell (2x_\ell^2+1)\sqrt{x_\ell^2+1} - {\rm ln}\left( x_\ell+\sqrt{x_\ell^2+1}\right) \right]
\end{equation}
where $\lambda_\ell = \hbar/m_\ell c$ is the Compton wave length and the dimensionless parameter $x_\ell$ is defined in terms 
of the Fermi wave number
\begin{equation}
k_{{\rm F}\ell}=  (3\pi^2 n_\ell)^{1/3}\, ,
\end{equation}
by
\begin{equation}
x_\ell=\lambda_\ell k_{{\rm F}\ell}\, .
\end{equation}
Since the electron mass is set to zero, the electron energy density is obtained by taking the limit $x_e \rightarrow +\infty$ of Eq.~(\ref{eq:lepton.energy}) yielding 
\begin{equation}
U_{\rm L}\left\{n_e\right\} = \frac{p_{{\rm F}e}^4}{4\pi^2(\hbar c)^3} \, ,
\end{equation}
where $p_{{\rm F}e}=\hbar k_{{\rm F}e}$  is the electron Fermi momentum. 

The strong interactions among nucleons are described by an effective Hamiltonian with a two-body force 
of the Skyrme type~\citep{bender-03,stone-07} 
\begin{eqnarray}
\label{eq:skyrme.force}
v\{\pmb{r_1},\pmb{r_2}\} = t_0 (1+x_0 P_\sigma) \delta\{\pmb{r}\} + \frac{t_1}{2}(1+x_1 P_\sigma)\left( \pmb{\hat k^\prime}^2\delta\{\pmb{r}\} + \delta\{\pmb{r}\} {\pmb{\hat k}}^2 \right) + t_2 (1+x_2 P_\sigma) \pmb{\hat k^\prime} \cdot \delta\{\pmb{r}\}\pmb{\hat k} \\
 + \frac{t_3}{6}(1+x_3 P_\sigma)\delta\{\pmb{r}\} n_{\rm b}\{\pmb{R}\}^\alpha + {\rm i} W_0 (\pmb{\hat \sigma_1} + \pmb{\hat \sigma_2})\cdot \pmb{\hat k^\prime} \times \delta\{\pmb{r}\}  \pmb{\hat k}
\end{eqnarray}
where $\pmb{r}=\pmb{r_1}-\pmb{r_2}$, $\pmb{R}=(\pmb{r_1}+\pmb{r_2})/2$, $\pmb{\hat \sigma_1}$ and $\pmb{\hat \sigma_2}$ are Pauli spin matrices, $\pmb{\hat k} = -{\rm i} (\pmb{\nabla_1} - \pmb{\nabla_2})/2$ is the relative wave vector, $\pmb{\hat k^\prime}$ is the complex conjugate of $\pmb{\hat k}$ acting on the left, and $P_\sigma=(1+\pmb{\hat \sigma_1}\cdot \pmb{\hat \sigma_2})/2$ is the spin-exchange operator. The first term represents the attractive part of the nucleon-nucleon interaction. The next two momentum dependent terms are associated with the finite range of the interaction. The density dependent term proportional to $t_3$ corresponds to the strongly repulsive short range part of the interaction and simulates the effects of three body interactions~\citep{vautherin-72}. The last term which leads to spin-orbit coupling in finite nuclei does not contribute in uniform matter. 
In principle, as shown by~\citet{nv-72}, this effective interaction can be derived from the ``bare'' nucleon-nucleon interaction by 
expanding the nucleon density matrix in relative and centre of mass coordinates, $\pmb{r}$, $\pmb{R}$ respectively. In practice however, the parameters are usually 
determined by fitting experimental data and/or results of microscopic many-body calculations in infinite uniform nuclear matter using the bare nucleon-nucleon interactions. Such kind of zero range 
effective forces are valid whenever the inter particle spacing is much larger than the range of the nuclear interactions. This condition 
is satisfied at densities below $\sim 3 \rho_0$. Nevertheless since these effective forces are usually constrained to reproduce the high 
density equation of state of nuclear matter, it is not completely unreasonable to apply them at densities above $\sim 3 \rho_0$. Finite range 
effects of the nucleon-nucleon interaction as well as relativistic corrections are somehow taken into account phenomenologically by the fitting 
procedure. For instance, the parametrizations SLy~\citep{chabanat-97} have been specifically constructed 
for neutron star studies by fitting a ``realistic'' equation of state of neutron matter up to very high densities $\sim 10 \rho_0$. Besides, it is worth mentioning that soon after such effective forces were introduced,~\citet{cameron-59} applied them to neutron stars and showed that the maximum mass $\sim 2 M_\odot$ is compatible with the scenario of neutron star formation from supernova explosions. 
The main limitation of these effective forces is that they describe only nucleonic degrees of freedom. At densities above $\sim 3 \rho_0$, other particles like hyperons are likely to appear~\citep*{haensel-06}. Nevertheless, let us remark that for the most precisely measured neutron star masses in binary radio pulsars, their central densities lie below $\sim 3-4 \rho_0$ depending on the equation of state (see for instance chapter 6 from ~\citet{haensel-06} and in particular their figure 6.3). Our main motivation for using such kind of effective Hamiltonian is the perspective of a unified treatment of the interior of neutron stars, including not only the liquid core but also the solid crust whose microscopic description starting from the bare nucleon-nucleon interaction is not feasible.

Assuming that the matter in neutron star cores is not polarized for the densities of interest as suggested by many-body calculations (see for instance~\citealt{bombaci-06}), the nucleon energy density associated with the force~(\ref{eq:skyrme.force}) can be calculated using the method outlined in the classic paper of \citet{vautherin-72} and is given by an expression of the form
\begin{equation}
U_{\rm N}\left\{n_n,n_p\right\}=n_{\rm b} m c^2+\frac{\hbar^2}{2 m}\tau_{\rm b} + B_1 n_{\rm b}^2 + B_2 (n_n^2+n_p^2) + B_3 n_{\rm b} \tau_{\rm b} + B_4 (n_n \tau_n+n_p \tau_p ) 
+ B_5 n_{\rm b}^{2+\alpha} + B_6 n_{\rm b}^\alpha (n_n^2+n_p^2) \, ,
\end{equation}
where 
\begin{equation}
\tau_n = \frac{3}{5}(3\pi^2)^{2/3} n_n^{5/3}\, , \hskip 0.5cm \tau_p = \frac{3}{5}(3\pi^2)^{2/3} n_p^{5/3} \, ,
\end{equation}
are respectively the neutron and proton kinetic energy densities (in units of $\hbar^2/2m$), and $\tau_{\rm b}=\tau_n+\tau_p$.
The $B$-coefficients are related to the parameters of the force by the following expressions
\begin{equation}
B_1 = \frac{t_0}{2}\left(1+\frac{x_0}{2}\right)
\end{equation}
\begin{equation}
B_2 = -\frac{t_0}{2}\left(x_0+\frac{1}{2}\right)
\end{equation}
\begin{equation}
B_3 = \frac{1}{4}\left[t_1\left(1+\frac{x_1}{2}\right) + t_2\left(1+\frac{x_2}{2}\right)\right]
\end{equation}
\begin{equation}
B_4 = -\frac{1}{4}\left[t_1\left(x_1+\frac{1}{2}\right) - t_2\left(x_2+\frac{1}{2}\right)\right]
\end{equation}
\begin{equation}
B_5 = \frac{t_3}{12}\left(1+\frac{x_3}{2}\right)
\end{equation}
\begin{equation}
B_6 = -\frac{t_3}{12}\left(x_3+\frac{1}{2}\right) \, .
\end{equation}

Analytic expressions of the effective masses $m_\star^n$ and $m_\star^p$ for the force~(\ref{eq:skyrme.force}) 
have been recently obtained by~\citet{chamelhaensel-06}. Introducing the parameter $\beta_3=2 m B_3/\hbar^2$, the coefficients 
of the mobility matrix, given by Eqs.~(\ref{eq:effmass}) and (\ref{eq.mobility.matrix}), can be 
expressed as 
\begin{equation}
\label{eq.kappann}
{\cal K}^{nn}=\frac{m}{n_n}\frac{1+\beta_3 n_n}{1+\beta_3 n_{\rm b}}
\end{equation}
\begin{equation}
\label{eq.kappapp}
{\cal K}^{pp}=\frac{m}{n_p}\frac{1+\beta_3 n_p}{1+\beta_3 n_{\rm b}}
\end{equation}
\begin{equation}
\label{eq.kappanp}
{\cal K}^{np}={\cal K}^{pn}= m \frac{\beta_3}{1+\beta_3 n_{\rm b}} \, .
\end{equation}
Note that as a consequence of the isospin symmetry of the nucleon-nucleon interactions, 
we have ${\cal K}^{nn}\{n_n,n_p\}={\cal K}^{pp}\{n_p,n_n\}$ and 
${\cal K}^{np}$ (therefore $\lambda_1$) does not depend on the matter 
composition but only on the \emph{total} baryon density $n_{\rm b}=n_n+n_p$. This means that entrainment effects are not affected by the 
various chemical reactions that may occur inside the core, as discussed in Section~\ref{sect.composition}. 
In the high density limit $\beta_3 n_n \gg 1$ and $\beta_3 n_p \gg 1$, all the elements of the mobility matrix 
become equal ${\cal K}^{q q^\prime} \rightarrow m/n_{\rm b}$. As a consequence, the non-relativistic effective masses 
tend to $m^q_\star/m \rightarrow n_q/n_{\rm b}$. This asymptotic limit which corresponds to the strongest entrainment effects
(see the discussion of Section~\ref{sect.non-rel.hydro}) is never reached in neutron star core for the nucleon-nucleon interactions
considered in this work, since $m/\beta_3$ is typically of the order of $\sim 10 \rho_0$ (see Table~\ref{table.forces.properties}).

\section{Choice of effective microscopic Hamiltonian}
\label{sect.examples}

We have selected effective forces according to the following criteria. First of all, the chosen forces have to yield reasonable values of 
the ``semi-empirical'' saturation properties of infinite uniform symmetric nuclear matter, namely the equilibrium or saturation density $n_0$ (or the mass density $\rho_0=n_0 m$), the binding energy per nucleon
\begin{equation}
a_v = \frac{U_{\rm N}\{n_0/2, n_0/2\} }{n_0} -m c^2 \, ,
\end{equation}
the symmetry energy coefficient
\begin{equation}
a_s = \frac{1}{2} \frac{\partial^2}{\partial I^2}  \left(\frac{U_{\rm N} }{n_{\rm b}}\right)\biggl\vert_{I=0, n_{\rm b}=n_0}
\end{equation}
with $I= (n_n-n_p)/n_{\rm b}$, and the incompressibility modulus 
\begin{equation}
K_\infty = 9 n_0^2 \frac{\partial^2}{\partial n_{\rm b}^2} \left(\frac{U_{\rm N}\{n_{\rm b}/2, n_{\rm b}/2\} }{n_{\rm b} }\right)\biggl\vert_{n_{\rm b}=n_0} \, .
\end{equation}
Global fits to essentially all the available experimental nuclear mass data yield $n_0 \simeq 0.16$ fm$^{-3}$, $a_v\simeq -16$ MeV, $a_s \simeq 28-35$ MeV and $K_\infty \simeq 220-240$ MeV~\citep{lunney-03}. 
Due to the strong interactions, the mass of the individual nucleons in nuclear matter is different from the bare mass and can be written as 
\begin{equation}
\frac{m}{m^*_n}=(1+I)\frac{m}{m^*_s} - I \frac{m}{m^*_v} \, , \hskip 0.5cm \frac{m}{m^*_p}=(1-I)\frac{m}{m^*_s} + I \frac{m}{m^*_v}
\end{equation}
in which $m^*_s$ and $m^*_v$ are the so-called isoscalar and isovector effective masses respectively (see for instance~\citealt{farine-01}). 
The isovector effective mass is a crucial microscopic input since it controls directly the strength of entrainment effects in neutron-proton 
mixtures. Indeed the parameter $\beta_3$ which determines the mobility matrix, Eqs.~(\ref{eq.kappann}),(\ref{eq.kappapp}) and (\ref{eq.kappanp}), is given by
\begin{equation}
n_{\rm b} \beta_3 =\frac{m}{m^{*}_v}-1 \, .
\end{equation} 
In principle, this isovector effective mass can be determined from measurements of the giant isovector electric dipole resonance in finite nuclei (consisting of relative motions between neutrons and protons). Nevertheless estimates are model dependent providing values $m^{*}_v/m\sim 0.7-1$ at saturation density (see in particular the discussion of ~\citealt{lunney-03} in Sect.~III-B-5-e). Microscopic many-body calculations in infinite uniform nuclear matter starting from the bare nucleon-nucleon interaction lead to an isovector effective mass around $m^{*}_v/m\sim 0.7$ (see for instance~\citealt{zuo-06}). Besides we consider only those effective forces that have been constrained to fit the uniform infinite neutron matter equation of state. Otherwise these effective forces could not be reliably extrapolated to the neutron rich matter inside neutron star core. 

The main deficiencies of effective forces is the existence of instabilities that are not found by microscopic calculations~\citep{margueron-02, agrawal-04,lesinski-06}. Especially many Skyrme forces predict a spurious ferromagnetic transition in neutron matter above some critical densities. 
We thus require that no such instabilities occur in the density range of interest $\rho < 3 \rho_0$ by imposing that the dimensionless Landau parameter, usually noted $G_0$, be greater than $-1$ in neutron matter (following the analysis of ~\citealt{margueron-02}). It turns out that 
this criterion is very restrictive. Several forces that reproduce reasonably well both the saturation properties of symmetric nuclear matter and the neutron matter equation of state do not pass this test. For instance, the parametrization RATP~\citep{ratp-82}, which was the first attempt to construct an effective force for astrophysical applications, predicts that neutron matter becomes spin polarized slightly above saturation density $\simeq 0.175$ fm$^{-3}$ (the density for the onset of instability is obtained by solving $G_0=-1$). Likewise the forces SkM and Skyrme $1^{\prime}$, which have been applied to study dense matter in neutron stars and supernova cores~\citep{bonche-82,lattimer-85,lassaut-87, lorenz-93}, yield a ferromagnetic transition density in neutron matter $\simeq 0.212$ fm$^{-3}$ and $\simeq 0.256$ fm$^{-3}$ respectively. We have found that only the forces of the Saclay-Lyon group~\citep{chabanat-97,chabanat-98,chabanat-err-98} and the recent parametrization LNS~\citep{cao-06} satisfy all the above conditions. They predict a ferromagnetic instability in neutron matter like the other forces, but at significantly higher densities $\sim 3-4 \rho_0$ which we do not consider in this work.
The force LNS seems the most appropriate to describe neutron star core since it was constructed so as to reproduce 
recent results of microscopic diagrammatic calculations (based on Brueckner theory) of infinite uniform nuclear matter with two- as well as three-body  forces. In particular, this effective force not only fits well the energy per nucleon in symmetric and asymmetric nuclear matter, but fits also 
the nucleon effective masses for different asymmetries and different densities which directly determine the entrainment coefficients as previously discussed. Nevertheless, the SLy forces, which were constrained to reproduce some properties of 
finite nuclei (apart from the other constraints that we imposed) would be preferable if not only the liquid core but also the crust layers
would have to be described with the same underlying microscopic Hamiltonian. Besides the equation of state of neutron star matter with the force SLy4 
has been tabulated and widely applied~\citep{haensel-04}. For comparison, we have also considered the parametrization NRAPR~\citep{steiner-05} since it was adjusted on the realistic equation of state of~\citet{akmal-98}. Nevertheless this force leads to a ferromagnetic instability at rather low density $\rho \lesssim 2 \rho_0$. 

\begin{table*}
 \centering
\caption{Parameters of the chosen Skyrme forces. The units of energy and length are MeV and fm respectively.}
\label{table.forces.parameters}
\begin{tabular}{|c c c c|}
\hline
& SLy4 & LNS & NRAPR\\
\hline
\hline
$t_0$ & -2488.91 & -2484.97 & -2719.7\\
$t_1$ & 486.82 & 266.735  & 417.64\\
$t_2$ & -546.39 & -337.135 & -66.687 \\
$t_3$ & 13777.0 & 14588.2 & 15042.0\\
$\alpha$ & $1/6$ & $1/6$ & 0.14416\\
$x_0$ & 0.834 & 0.06277 & 0.16154\\
$x_1$ & -0.344 & 0.65845 & -0.0047986\\
$x_2$ & -1 & -0.95382  & 0.027170\\
$x_3$ & 1.354 & -0.03413 & 0.13611\\
\hline
\end{tabular}
\end{table*}

\begin{table*}
\centering
\caption{B-coefficients of the chosen Skyrme forces. The units of energy and length are MeV and fm respectively.}
\label{table.forces.Bcoef}
\begin{tabular}{|c c c c c c c|}
\hline
& $B_1$ & $B_2$ & $B_3$ & $B_4$ & $B_5$ & $B_6$ \\
\hline
\hline
SLy4 & -1763.39 & 1660.1 & 32.473 & 49.3128 & 1925.34 & -2128.55  \\
LNS & -1281.48 & 699.233 & 44.5497 & -39.0001 & 1194.94 & -566.35 \\
NRAPR & -1469.69 & 899.595 & 85.0067 & -55.9836 & 1338.81 & -797.364 \\
\hline
\end{tabular}
\end{table*}

\begin{figure}
\begin{center}
\epsfig{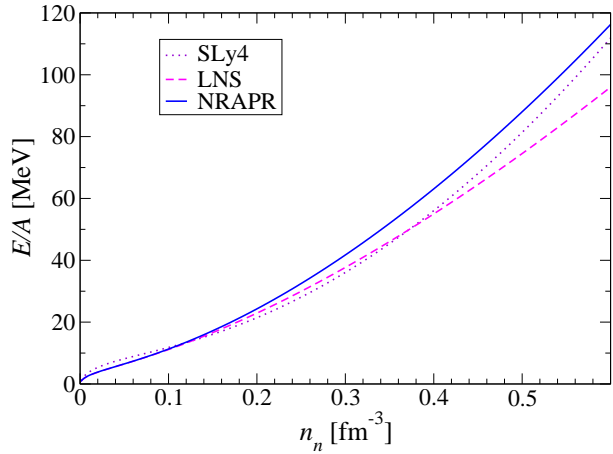}
\end{center}
\caption{Binding energy per particle of uniform infinite neutron matter for the SLy4, LNS and NRAPR effective forces.}
\label{fig:eos_pnm}
\end{figure}

The parameters of the forces and the associated B-coefficients introduced in Section~\ref{sect.micro}, 
are given in Tables~\ref{table.forces.parameters} and ~\ref{table.forces.Bcoef} respectively. The nuclear 
matter properties predicted by these forces are summarized in Table~\ref{table.forces.properties}. Figure~\ref{fig:eos_pnm}
shows the binding energy per particle in uniform infinite neutron matter defined by $E/A=U_{\rm N}\{n_n,0\}/n_n -m c^2$. 
Let us stress that the LNS force was fitted to the latest results of 
many body calculations with two- and three-body forces, while the forces of the Lyon group were adjusted to 
reproduce an older neutron matter equation of state based on variational methods. 

\begin{figure}
\begin{center}
\epsfig{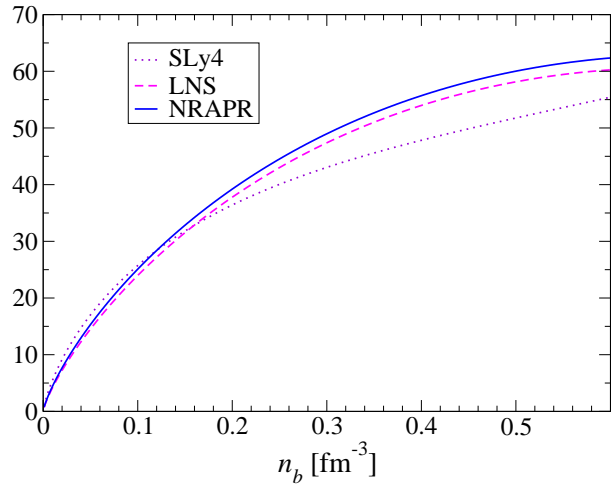}
\end{center}
\caption{Symmetry energy ${\cal S}\{n_{\rm b}\}$ (in MeV) for the SLy4, LNS and NRAPR effective forces as a function of the baryon density $n_{\rm b}=n_n+n_p$.}
\label{fig:sym}
\end{figure}

\begin{figure}
\begin{center}
\epsfig{file=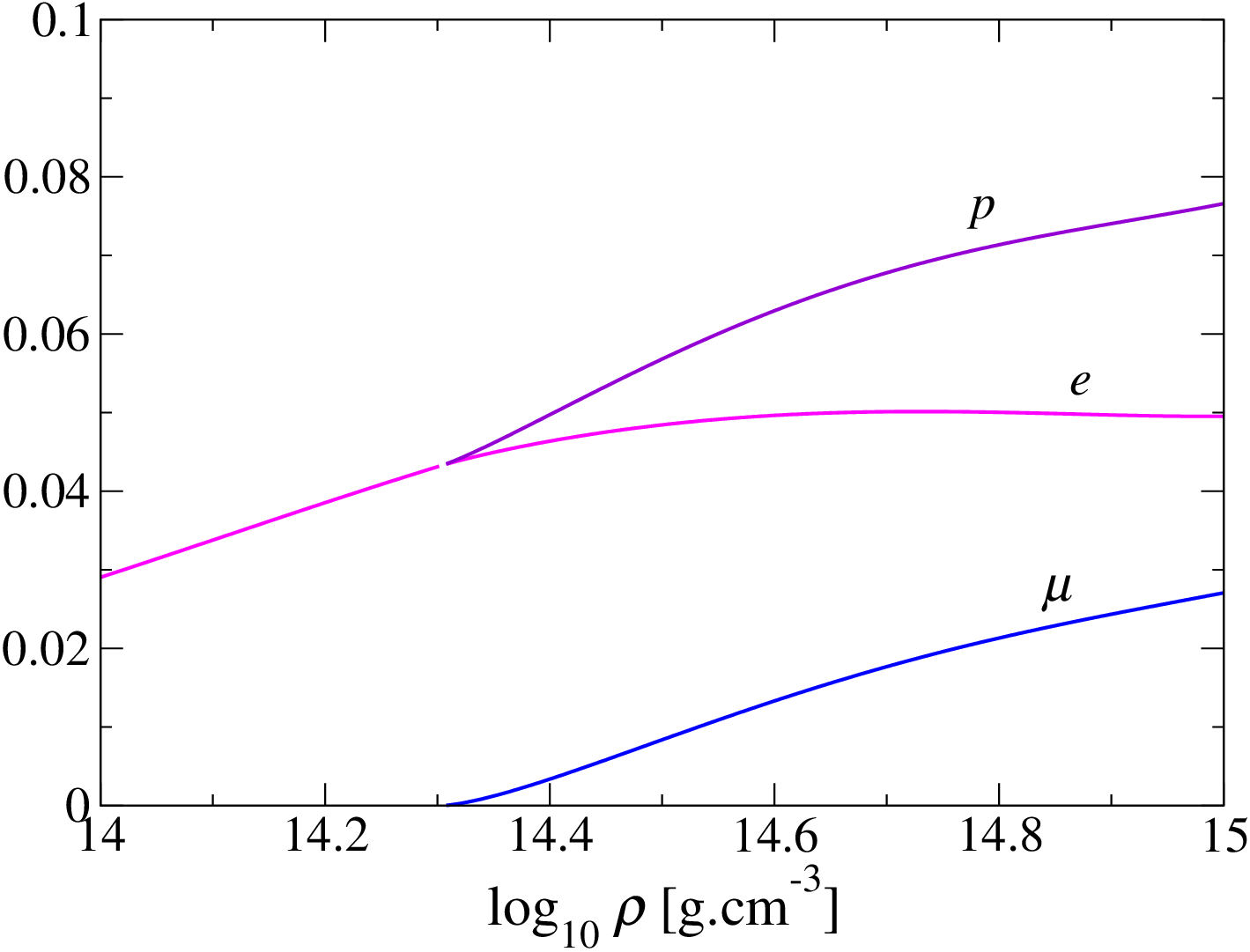,width=8.cm}
\end{center}
\caption{Equilibrium fractions $n_{_{\rm X}}/n_{\rm b}$ of protons (p), electrons (e) and muons ($\mu$) inside neutron star core predicted by the SLy4 effective force.}
\label{fig:SLy4_core}
\end{figure}

\begin{figure}
\begin{center}
\epsfig{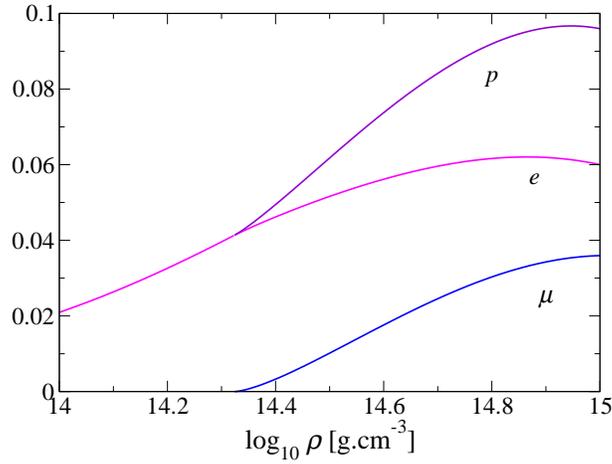}
\end{center}
\caption{Equilibrium fractions $n_{_{\rm X}}/n_{\rm b}$ of protons (p), electrons (e) and muons ($\mu$) inside neutron star core predicted by the LNS effective force.}
\label{fig:LNS_core}
\end{figure}

\begin{figure}
\begin{center}
\epsfig{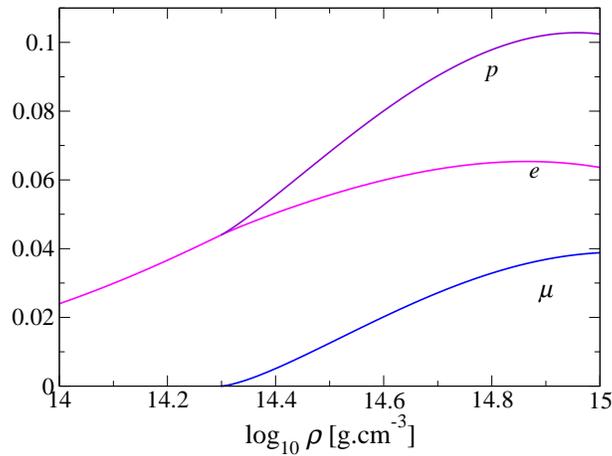}
\end{center}
\caption{Equilibrium fractions $n_{_{\rm X}}/n_{\rm b}$ of protons (p), electrons (e) and muons ($\mu$) inside neutron star core predicted by the NRAPR effective force.}
\label{fig:NRAPR_core}
\end{figure}

Figures~\ref{fig:SLy4_core}, \ref{fig:LNS_core} and \ref{fig:NRAPR_core} show the equilibrium composition of cold neutron star matter, composed of neutrons, protons, electrons and muons, obtained by solving Eqs.~(\ref{eq.neutrality}), (\ref{eq.beta.equi}) and (\ref{eq.beta.equi2}). The figures show the electron, muon and proton fractions, defined respectively by $n_e/n_{\rm b}$, $n_\mu/n_{\rm b}$ and $n_p/n_{\rm b}$, as a function of the mass-energy density $\rho=U_{\rm ins}/c^2$, which is approximately given by $\rho\simeq n_{\rm b} m$ for $n_{\rm b} < 3 n_0$. All forces predict the appearance of muons at $n_{\rm b}\simeq0.12$ fm$^{-3}$ or $\rho\simeq 2\times 10^{14}$ g.cm$^{-3}$). 
The forces LNS and NRAPR yield similar composition. They both predict a slightly larger (resp. smaller) proton fraction than the force SLy4 above (resp. below) $\rho_0$. This can be understood by remarking that Eq.~(\ref{eq.beta.equi})
can be approximately written as~\citep{muether-87}
\begin{equation}
\label{eq.muther}
\hbar c(3\pi^2 n_{\rm b} x_e)^{1/3} \approx 4 {\cal S}\{n_{\rm b}\} (1-2 x_p) \, ,
\end{equation}
where the symmetry energy ${\cal S}\{n_{\rm b}\}$ defined by 
\begin{equation}
{\cal S}\{n_{\rm b}\} = \frac{U_{\rm N}\{n_{\rm b},0\}}{n_{\rm b}} - \frac{U_{\rm N}\{n_{\rm b}/2,n_{\rm b}/2\}}{n_{\rm b}} >0 \, ,
\end{equation}
represents the cost in (nuclear) energy per particle to replace protons by neutrons in symmetric nuclear matter.
From Eq.~(\ref{eq.muther}), we have 
\begin{equation}
x_p \approx x_e \approx \left(\frac{4 {\cal S}\{n_{\rm b}\}}{\hbar c}\right)^3 \frac{1}{3\pi^2 n_{\rm b}} \, .
\end{equation}
As can be seen on Figure~\ref{fig:sym}, the forces LNS and NRAPR yield a larger (resp. smaller) symmetry energy than the force SLy4 above (resp. below) the saturation density $n_0$ (note that the symmetry energy coefficient $a_s\simeq {\cal S}\{n_0\}$). 

\begin{figure}
\begin{center}
\epsfig{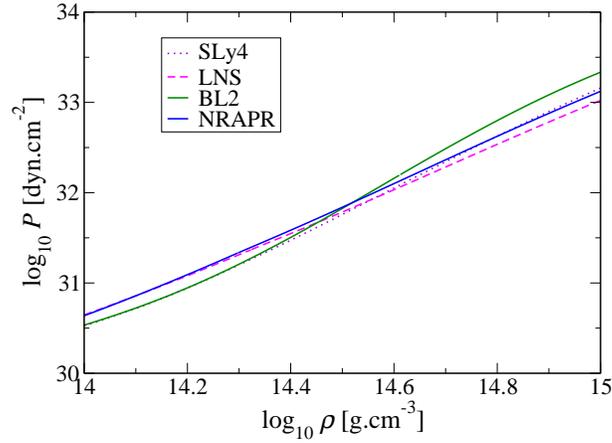}
\end{center}
\caption{Static pressure of neutron star matter in equilibrium for the Skyrme effective forces and for the BL2 relativistic mean field model.}
\label{fig:eos}
\end{figure}

\begin{figure}
\begin{center}
\epsfig{file=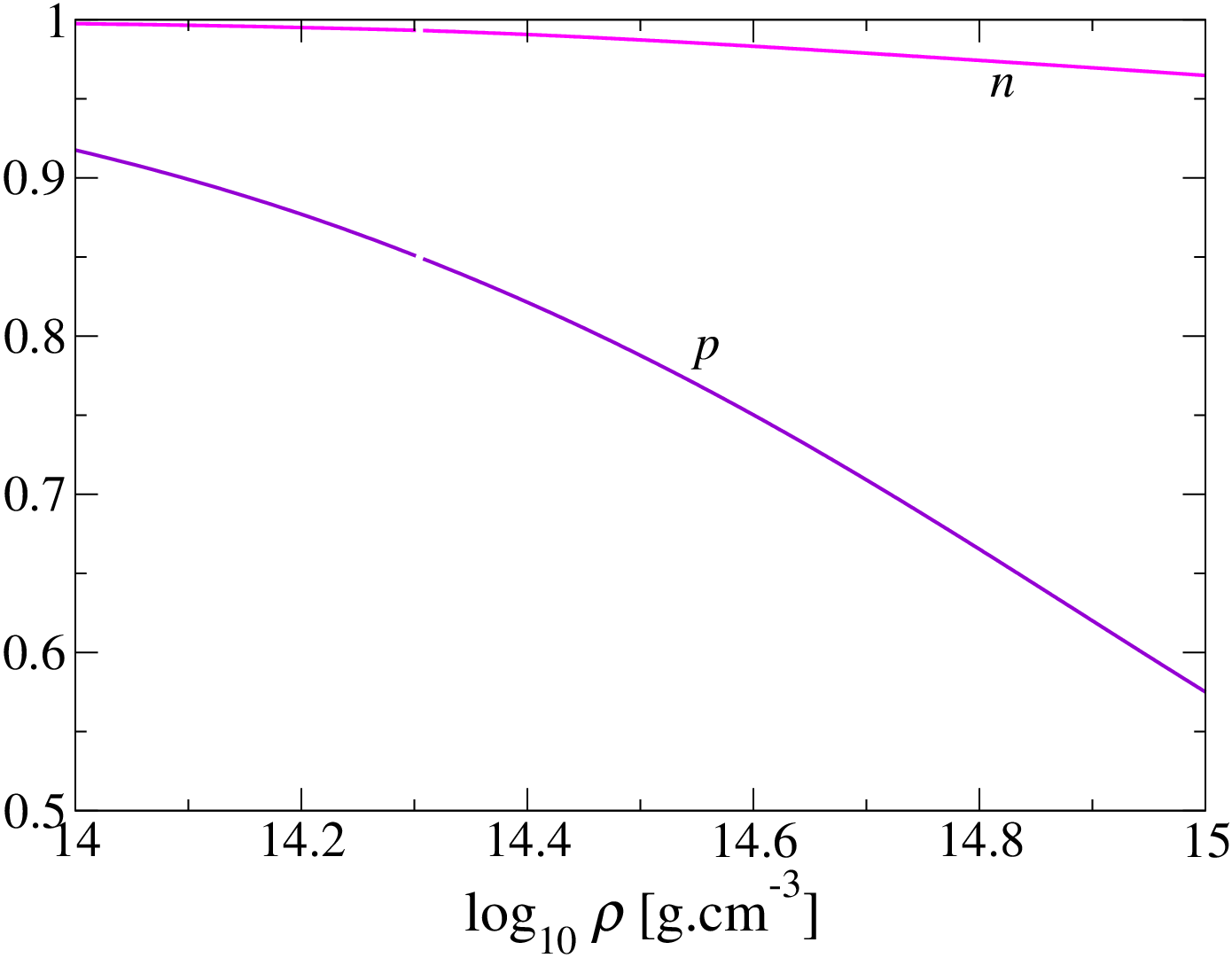,width=8.cm}
\end{center}
\caption{Neutron and proton effective masses, respectively $m^n_\star/m$ and $m^p_\star/m$, defined by Eq.~(\ref{eq:effmass}), in neutron star matter in equilibrium for the SLy4 effective force. }
\label{fig:effmass_SLy4}
\end{figure}

\begin{figure}
\begin{center}
\epsfig{file=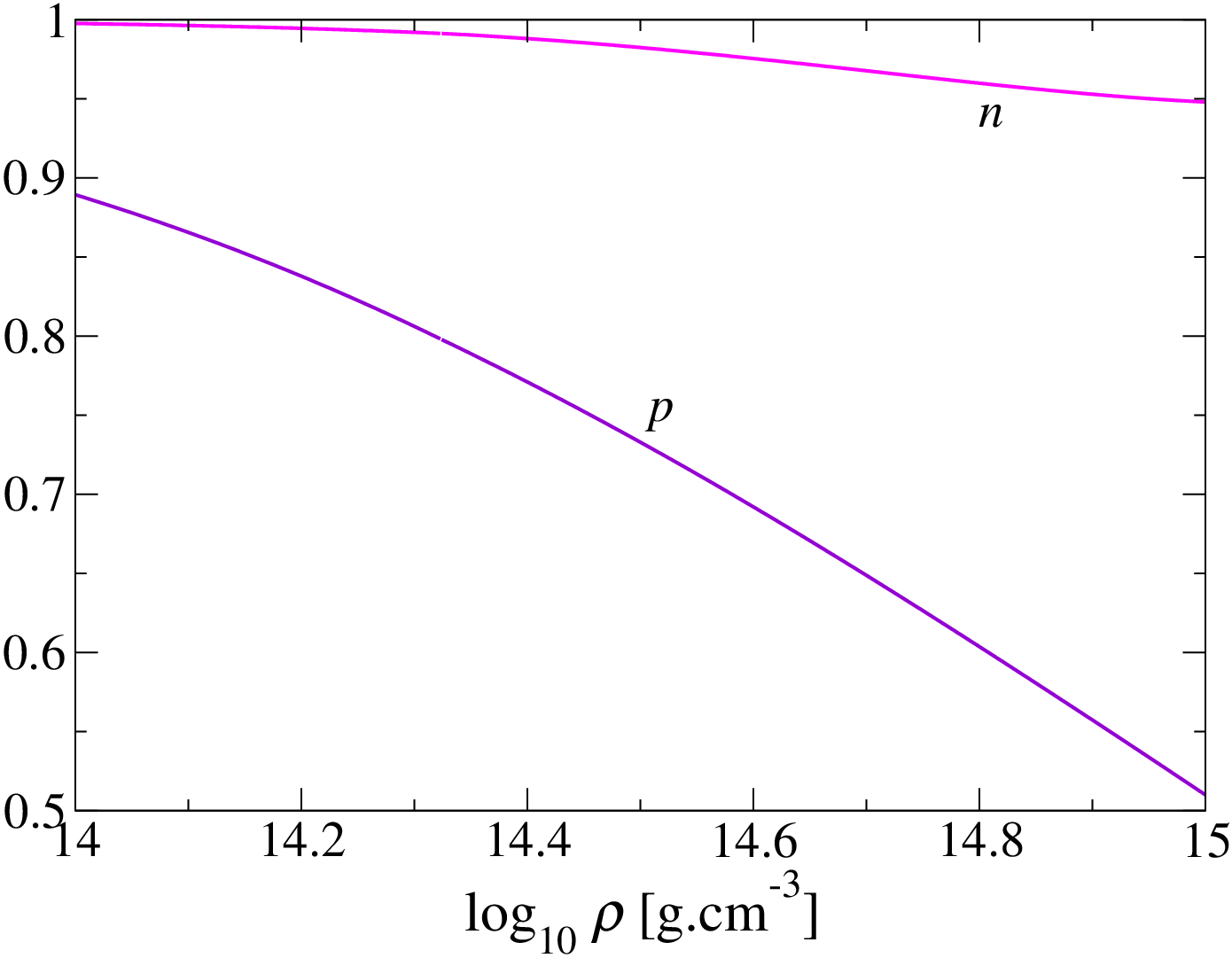,width=8.cm}
\end{center}
\caption{Neutron and proton effective masses, respectively $m^n_\star/m$ and $m^p_\star/m$, defined by Eq.~(\ref{eq:effmass}), in neutron star matter in equilibrium for the LNS effective force. }
\label{fig:effmass_LNS}
\end{figure}

\begin{figure}
\begin{center}
\epsfig{file=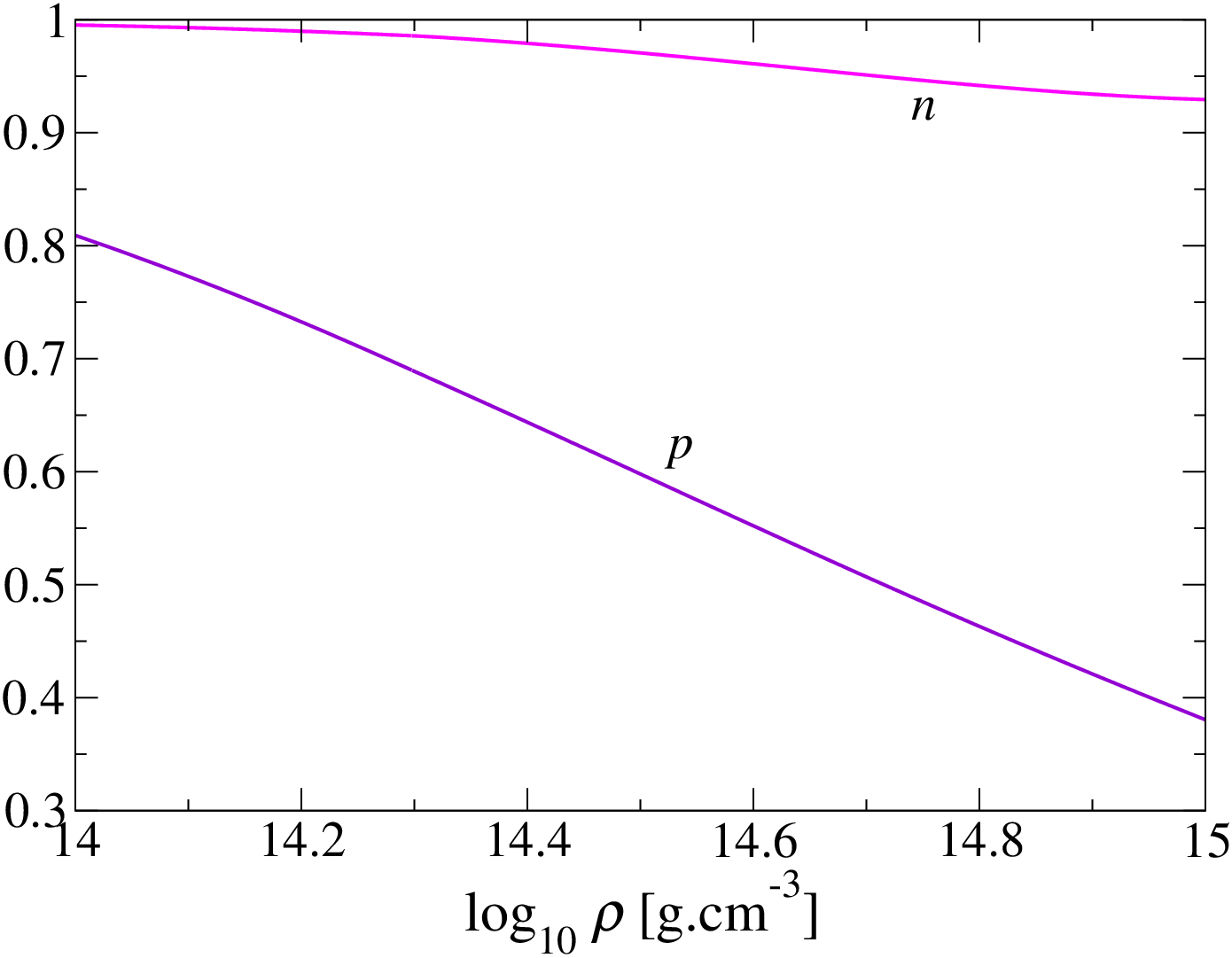,width=8.cm}
\end{center}
\caption{Neutron and proton effective masses, respectively $m^n_\star/m$ and $m^p_\star/m$, defined by Eq.~(\ref{eq:effmass}), in neutron star matter in equilibrium for the NRAPR effective force. }
\label{fig:effmass_NRAPR}
\end{figure}

\begin{figure}
\begin{center}
\epsfig{file=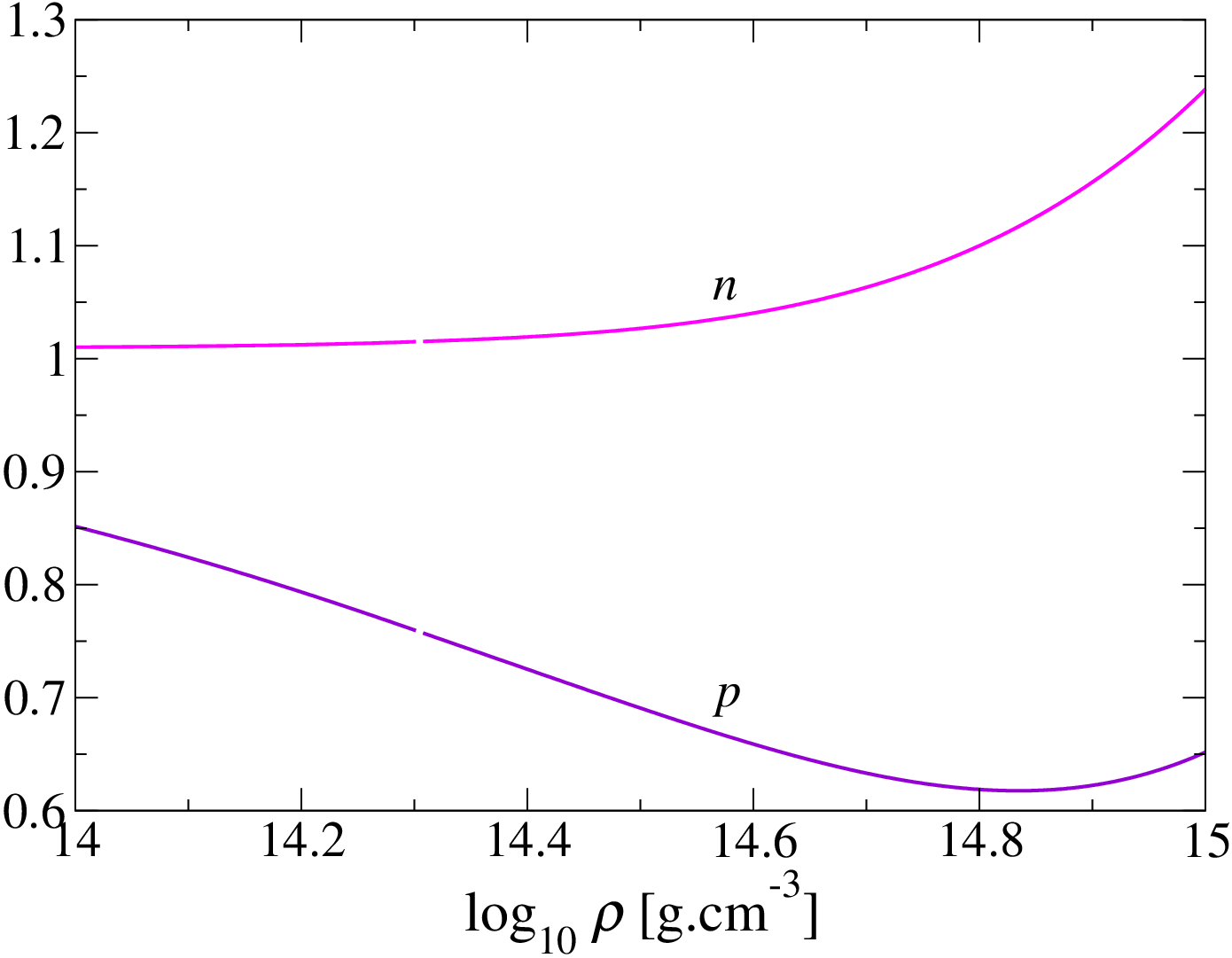,width=8.cm}
\end{center}
\caption{Relativistic neutron and proton effective masses, respectively $\widetilde{m}^n_\star/m$ and $\widetilde{m}^p_\star/m$, defined by Eq.~(\ref{eq:releffmass}), in neutron star matter in equilibrium for the SLy4 effective force. Neutrons and protons are co-moving. }
\label{fig:releffmass_SLy4}
\end{figure}

\begin{figure}
\begin{center}
\epsfig{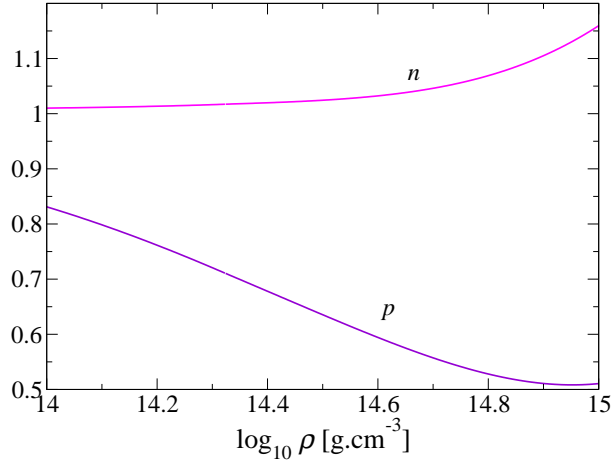}
\end{center}
\caption{Relativistic neutron and proton effective masses, respectively $\widetilde{m}^n_\star/m$ and $\widetilde{m}^p_\star/m$, defined by Eq.~(\ref{eq:releffmass}), in neutron star matter in equilibrium for the LNS effective force. Neutrons and protons are co-moving. }
\label{fig:releffmass_LNS}
\end{figure}

\begin{figure}
\begin{center}
\epsfig{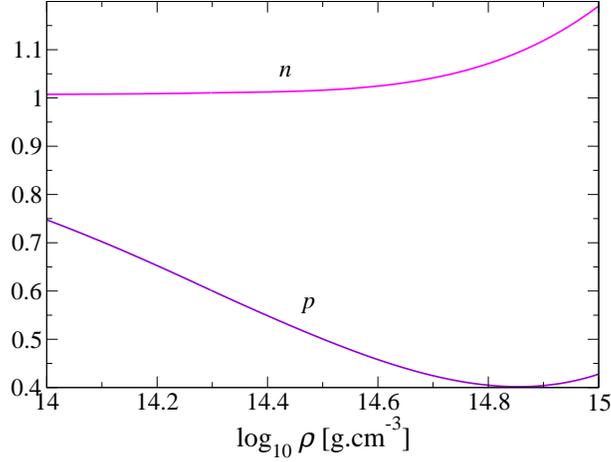}
\end{center}
\caption{Relativistic neutron and proton effective masses, respectively $\widetilde{m}^n_\star/m$ and $\widetilde{m}^p_\star/m$, defined by Eq.~(\ref{eq:releffmass}), in neutron star matter in equilibrium for the NRAPR effective force. Neutrons and protons are co-moving. }
\label{fig:releffmass_NRAPR}
\end{figure}

Figure~\ref{fig:eos} shows the static pressure~(\ref{eq.Peq}) of $npe\mu$ matter in equilibrium as a function of the mass-energy density 
$\rho=U_{\rm ins}/c^2$. In figures~\ref{fig:effmass_SLy4}, \ref{fig:effmass_LNS} and \ref{fig:effmass_NRAPR}, we compare the effective masses defined by Eq.~(\ref{eq:effmass}) for the two Skyrme forces. In both cases the neutron effective mass is close to the bare nucleon mass while the proton effective mass is significantly reduced. This is consistent with the inequality~(\ref{eq.stability}) which implies that 
\begin{equation}
\label{eq.effmass.split}
\frac{m_\star^n}{m} - \frac{m_\star^p}{m} > I \, ,
\end{equation}
with $I= (n_n-n_p)/n_{\rm b}$. 
The relativistic effective masses defined by Eq.~(\ref{eq:releffmass}) are shown on Figures~\ref{fig:releffmass_SLy4}, \ref{fig:releffmass_LNS} and \ref{fig:releffmass_NRAPR}. For simplicity we have considered that neutrons and protons are co-moving so that 
the effective masses are given by
 \begin{equation}
\frac{\widetilde{m}_\star^q}{m} = \frac{m_\star^q}{m} + \frac{\mu_q}{m c^2} - 1  \, .
\end{equation}
These relativistic effective masses are significantly different compared to the non-relativistic ones. In particular both nuclear forces 
predict that the neutron relativistic effective mass is larger than the bare mass. Moreover, from Eqs.~(\ref{eq.effmass.split}) and (\ref{eq.beta.equi}), we have 
\begin{equation}
\frac{\widetilde{m}_\star^n}{m} - \frac{\widetilde{m}_\star^p}{m} = \frac{m_\star^n}{m} - \frac{m_\star^p}{m} + \frac{\mu_e}{m c^2} > I+\frac{\mu_e}{m c^2} \, ,
\end{equation}
so that the splitting of the relativistic effective masses is larger than that of the non-relativistic ones since $\mu_e\geq 0$. 

\begin{table*}
\centering
\caption{Properties of infinite uniform symmetric nuclear matter for the chosen Skyrme forces. $n_0$ (fm$^{-3}$) is the nuclear saturation density, $a_v$ (MeV) is the binding energy per nucleon of infinite symmetric nuclear matter, $a_s$ (MeV) the symmetry energy coefficient, $K_\infty$ (MeV) the compression modulus, $m^{*}_s/m$ and $m^{*}_v/m$ are the isoscalar and isovector effective masses respectively, $n_{\rm f}$ (fm$^{-3}$) is the density at which a ferromagnetic instability occurs in neutron matter. Realistic values of these parameters are 
$n_0\simeq 0.16$ fm$^{-3}$, $a_v\simeq -16$ MeV, $a_s\simeq 28-35$ MeV, $K_\infty\simeq220-240$ MeV, $m^{*}_s/m\sim 0.6-0.9$, $m^{*}_v/m\sim 0.7-1$~\citep{lunney-03}. }
\label{table.forces.properties}
\begin{tabular}{|c c c c c c c c|}
\hline
& $n_0$ & $a_v$ & $a_s$ & $K_\infty$ & $m^{*}_s/m$ &  $m^{*}_v/m$ & $n_{\rm f}$\\
\hline
\hline
SLy4 & 0.160 & -15.97 & 32.0 & 229.9 & 0.696 & 0.801 & 0.59 \\
LNS & 0.175 & -15.32 & 33.4 & 210.9 & 0.827 & 0.728 & 0.62\\
NRAPR & 0.1606 & -15.86 & 32.79 & 225.7 & 0.695 & 0.605 & 0.28 \\
\hline
\end{tabular}
\end{table*}

\section{Comparison with relativistic mean field models}
\label{sect.rmf}

A few years ago, \citet{comer-03}  developed relativistic two-fluid models of superfluid neutron star cores. 
They have determined the master function $\Lambda$ using the effective relativistic mean field theory~\citep{glendenning-00}. 
In the model they considered, the interactions between nucleons arise from the exchange of two massive mesons: the scalar meson $\sigma$ with mass $m_\sigma$ and the vector meson $\omega_\mu$ with mass $m_\omega$. The former accounts for the long range attractive part of the nucleon-nucleon interaction while the latter gives rise to the short range repulsive part. In the nuclear field theory, particles are described by a microscopic Lagrangian density $\cal L$ (not to be confused with the macroscopic Lagrangian density $\Lambda$ of the fluids) given by (using units $c=\hbar=1$)
\begin{equation}
\label{eq.rmf.lagrangian}
{\cal L} = \bar \psi [\gamma^\mu {\rm i} D_\mu - m_D]\psi + \frac{1}{2} [(\partial^\mu\sigma)( \partial_\mu \sigma) - m_\sigma^2 \sigma^2] - \frac{1}{4}\omega^{\mu\nu}\omega_{\mu\nu}+\frac{1}{2} m_\omega^2 \omega_\mu\omega^\mu
\end{equation}
with the nucleon field $\psi$ and the antisymmetric tensor $\omega_{\mu\nu}\equiv\partial_\mu \omega_\nu - \partial_\nu \omega_\mu$ ($\gamma^\mu$ denote the Dirac 
matrices and $\bar\psi\equiv\psi^\dagger \gamma^0$). The nucleon-meson couplings are introduced in the gauge covariant derivative 
\begin{equation}
D_\mu = \partial_\mu + {\rm i}\, g_\omega \omega_\mu
\end{equation}
and in the Dirac effective nucleon mass
\begin{equation}
m_D = m- g_\sigma \sigma
\end{equation}
where $g_\omega$ and $g_\sigma$ are dimensionless coupling constants. 
The field equations, which actually only depend on the quantities $c_\sigma^2=g_\sigma^2/m_\sigma^2$ and $c_\omega^2=g_\sigma^2/m_\sigma^2$ 
in uniform infinite matter, are solved in the Hartree approximation (exchange terms are neglected) ignoring the contributions of antiparticles
(the so-called no Dirac sea approximation). 

\begin{figure}
\begin{center}
\epsfig{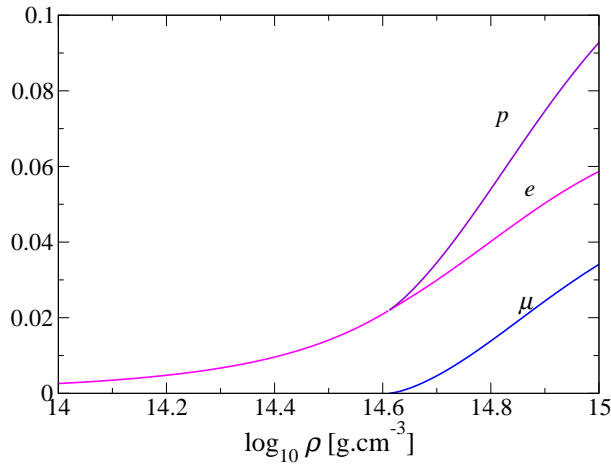}
\end{center}
\caption{Equilibrium fractions $n_{_{\rm X}}/n_{\rm b}$ of protons (p), electrons (e) and muons ($\mu$) inside neutron star core predicted by the BL2 relativistic mean field model.}
\label{fig:BL2_core}
\end{figure}

\begin{figure}
\begin{center}
\epsfig{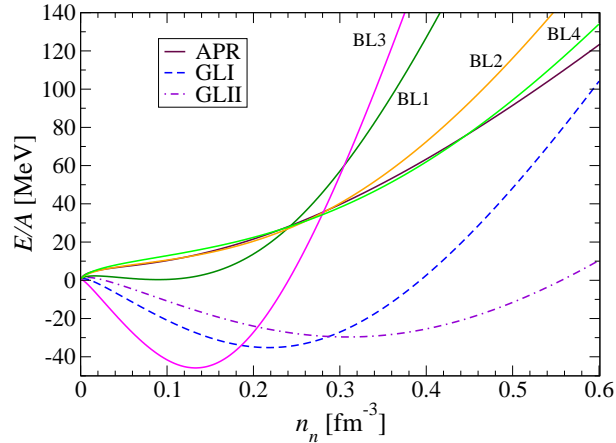}
\end{center}
\caption{Energy per particle of uniform infinite neutron matter for relativistic $\sigma-\omega$ mean field models. The curve labelled APR is 
the ``realistic'' equation of state of~\citet{akmal-98}, using the analytical fit of~\citet{heiselberg-00}. The rest mass energy
has been subtracted out. GLI and GLII are the models used by \citet{comer-03}. BL1-BL4 are the parameter set 
constructed in this work.}
\label{fig:eos_pnm_rmf}
\end{figure}

As for the non-relativistic energy density functional theory discussed in Section~\ref{sect.examples}, 
the free parameters of the model have to be determined by fitting to some nuclear matter properties. 
\citet{comer-03} adopted two parameter sets from \citet{glendenning-00} (chapter 4, table 4.4). However these parameters were not obtained for the 
Lagrangian density given by Eq.~(\ref{eq.rmf.lagrangian}) but for a more elaborate class of models which includes scalar self-interactions 
as well as the vector-isovector rho meson. As a result, by dropping these extra terms in the Lagrangian density $\cal L$, the resulting $\sigma-\omega$ models considered by \citet{comer-03} predict unphysical nuclear matter properties, as can
be seen in Table~\ref{table.rmf.properties} (see the discussion in Section~\ref{sect.examples}). Besides these models predict that neutron matter 
is bound as shown in Figure~\ref{fig:eos_pnm_rmf}, unlike quantum many body calculations using realistic nucleon-nucleon interactions. Note also 
that these models predict that neutrons and protons have the same (Dirac) effective mass $m_D$ for any nuclear asymmetry $I= (n_n-n_p)/n_{\rm b}$, in contradiction to microscopic calculations (see for instance~\citealt{vandalen-07}, especially their figure 4). As such, these models are therefore unsuitable for applications to neutron stars as pointed out by \citet{glendenning-00}. In order to reproduce the properties of finite nuclei and infinite nuclear matter with the same level of accuracy as the non-relativistic effective forces discussed in Section~\ref{sect.examples}, other mesons must be included. Besides non-linear self-meson interactions must be introduced (see for instance chapter 4 of~\citealt{glendenning-00}). 

For the present time, we will restrict the discussion of the entrainment effects to the $\sigma-\omega$ model. 
The models considered by \citet{comer-03} can be significantly improved (keeping in mind the inherent limitations of such models) 
by simply refitting the parameters. With only two free parameters $c_\sigma$ and $c_\omega$, only two of the symmetric 
infinite nuclear matter properties listed in Table~\ref{table.forces.properties} can be fitted exactly. We have constructed 
three new parameter sets BL1-BL3 by fixing the saturation density to $n_0=0.16$ fm$^{-3}$ and (i) the binding energy per 
nucleon to $a_v=-16$ MeV for BL1, (ii) the Dirac effective mass to $m_D=0.7 m$ for BL2, (iii) the symmetry energy to $a_s=28$ MeV 
for BL3 (the fitting procedure did not converge for $a_s=30$ MeV). The parameters of these new models are given in 
Table~\ref{table.meson.couplings}. As can be seen in comparing 
Table~\ref{table.rmf.properties} and \ref{table.forces.properties}, the overall agreement with empirical nuclear data is 
still very poor reflecting the lack of flexibility of these $\sigma-\omega$ models. The most important constraint for 
application to neutron stars is to reproduce at least the equation of state of neutron matter. We have thus constructed 
the parameter set BL4 by fitting the realistic equation of state of \citet{akmal-98}. The result of the fit is shown in 
Figure~\ref{fig:eos_pnm_rmf}, as well as the predictions of the other mean field models. This should be compared with the 
results of non-relativistic effective forces in Figure~\ref{fig:eos_pnm}. Note that the parameter sets GLI, GLII and 
BL3 predict incorrectly the existence of bound neutron matter. 
From these four models, it seems that the best compromise is achieved for the parameter set BL2, yielding reasonable values of the 
saturation density, compression modulus, Dirac effective mass (which is an important quantity for entrainment effects as discussed 
in Section~\ref{sect.examples}) together with a fairly good fit of the neutron matter equation of state. 
Note however that this model is still very crude compared to the SLy4 or LNS Skyrme forces presented in Section~\ref{sect.examples}. 
In particular, the values of the symmetry energy $a_s$ and the binding energy per nucleon  $a_v$, which are two basic nuclear matter properties, 
are unrealistic.

We have applied the general expressions derived by \citet{comer-03} within the $\sigma-\omega$ mean field models, to evaluate the entrainment 
parameters and to compare the results with those obtained using the non-relativistic effective energy density functional theory. 
We have greatly improved the models considered by \citet{comer-03} (i) by refitting the meson coupling constants leading to a better 
agreement with nuclear data as discussed previously, and (ii) by including muons which affect the composition of neutron star core and 
contribute to the pressure.
Leptons are treated as ideal relativistic Fermi gases as discussed in Section~\ref{sect.examples}. The parameters $\lambda_0$ and 
$\lambda_1$ in the expansion of the master function $\Lambda$, introduced in Section~\ref{sect.rel.hydro}, are given by
\begin{equation}
\lambda_0 = \Lambda\vert_0 - U_{\rm L}\left\{n_\mu\right\} \, , \hskip1cm \lambda_1=-{\cal A}\vert_0
\end{equation}
where ${\cal A}\vert_0$ and $\Lambda\vert_0$ are given by Eqs.(63) and (A8) respectively in the paper of \citet{comer-03}. Note that 
we have added the muon contribution in the Lagrangian density. Using the nucleon chemical potentials given by Eqs.(A9) and (A10) of that paper, 
together with the lepton chemical potentials defined by Eq.(\ref{eq.muX}), we have determined the equilibrium composition of the 
neutron star core assuming co-moving particles as discussed in Section~\ref{sect.composition}. 
Results are shown in Figure~\ref{fig:BL2_core} for the 
parameter set BL2. 
The proton fraction is very small at low densities unlike that predicted by non-relativistic effective forces. 
As discussed in Section~\ref{sect.examples}, this can be understood from the very small (incorrect) value of the symmetry energy $a_s$ at saturation density (see Table~\ref{table.rmf.properties}). 
As can be seen in Figure~\ref{fig:eos}, the equation of state is however similar to that 
obtained for the non-relativistic effective nucleon-nucleon interactions. The reason is that matter in neutron star core 
is almost pure neutron matter and all models SLy4, LNS and BL2 reproduce reasonably well the neutron matter equation of state 
(see Figure~\ref{fig:eos_pnm} and \ref{fig:eos_pnm_rmf}). 

\begin{figure}
\begin{center}
\epsfig{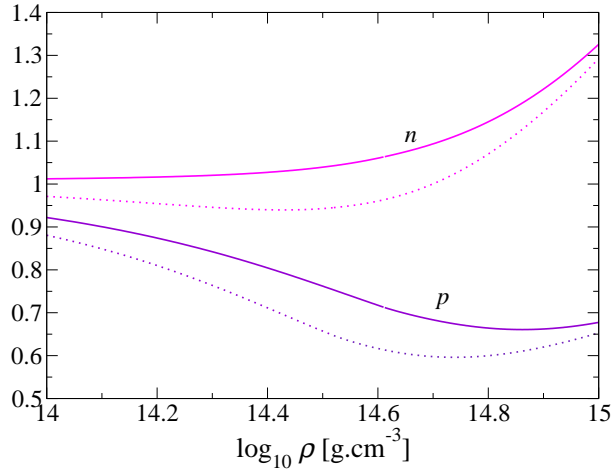}
\end{center}
\caption{Relativistic neutron and proton effective masses, respectively $\widetilde{m}^n_\star/m$ and $\widetilde{m}^p_\star/m$, defined by Eq.~(\ref{eq:releffmass}), in neutron star matter in equilibrium for the BL2 (solid lines) and GLI (dotted lines) relativistic mean field 
models. Neutrons and protons are co-moving. }
\label{fig:releffmass_BL2}
\end{figure}

The relativistic effective masses introduced in Section~\ref{sect.rel.hydro}
are given by 
\begin{equation}
\widetilde{m}^n_\star = n_n\, {\cal B}\vert_0 \, , \hskip1cm \widetilde{m}^p_\star=n_p\, {\cal C}\vert_0 
\end{equation}
where ${\cal B}\vert_0$ and ${\cal C}\vert_0$ are given by Eq.~(64) and (65) respectively in the paper of \citet{comer-03}. As shown in Figure~\ref{fig:releffmass_BL2}, the neutron effective mass predicted by the model GL2 is larger than the bare nucleon mass while the proton 
effective mass is smaller, as obtained for non-relativistic models. For comparison we have also plotted the effective masses obtained with 
the model GLI considered by \citet{comer-03}. The neutron effective mass obtained with this parameter set is decreased at low densities 
compared to the ordinary mass in contradiction to previous results. This is a consequence 
of the fact that the model GLI predicts (incorrectly) that neutron matter is bound, as shown in Figure~\ref{fig:eos_pnm_rmf}, so 
that $\mu_n < m c^2$.

\begin{table*}
\centering
\caption{Parameters of the $\sigma-\omega$ relativistic mean field models discussed in this work. GLI and GLII are the models proposed by 
\citet{glendenning-00} and employed by \citet{comer-03} after dropping scalar self-interactions and the rho meson. The models BL1-BL4 are new parameter sets introduced in this work (the coupling constants are given in fm$^2$). Note that for the baryon mass we have taken the average of the neutron and proton masses as \citet{comer-03}.}
\label{table.meson.couplings}
\begin{tabular}{|c c c|}
\hline
& $c_\sigma^2$ & $c_\omega^2$  \\
\hline
\hline
GLI & 12.684 & 7.148  \\
GLII & 8.403 & 4.233  \\
BL1 & 14.6063 & 11.0544 \\
BL2 & 9.3353 & 7.2624 \\
BL3 & 23.2707 & 14.4061 \\
BL4 & 7.79346 & 6.06748 \\
\hline
\end{tabular}
\end{table*}

\begin{table*}
\centering
\caption{Properties of infinite uniform symmetric nuclear matter for $\sigma-\omega$ relativistic mean field models. The quantities shown
are the same as those introduced in Table~\ref{table.forces.properties}. Realistic values are 
$n_0\simeq 0.16$ fm$^{-3}$, $a_v\simeq -16$ MeV, $a_s\simeq 28-35$ MeV, $K_\infty\simeq220-240$ MeV~\citep{lunney-03}. 
Note that recent relativistic many body calculations by~\citet{vandalen-07} predict that the Dirac effective mass at saturation is $m_D\simeq0.7 m$. GLI and GLII are the models employed by \citet{comer-03} while 
models BL1-BL4 are new parameter sets introduced in this work.}
\label{table.rmf.properties}
\begin{tabular}{|c c c c c c|}
\hline
& $n_0$ & $a_v$ & $a_s$ & $K_\infty$ & $m_D/m$\\
\hline
\hline
GLI & 0.28 & -66.43 & 34.69 & 2117.5 & 0.385   \\
GLII & 0.41 & -66.69 & 41.29 & 2252.0 & 0.407  \\
BL1 & 0.16 & -16 & 20.09 & 674.0 & 0.543  \\
BL2 & 0.16 & -1.70 & 16.29 & 249.0 & 0.700  \\
BL3 & 0.16 & -72.24 & 28 & 2217.4 & 0.338  \\
BL4 & 0.14 & 1.7358 & 13.88 & 130.5 & 0.774  \\
\hline
\end{tabular}
\end{table*}

\section{Conclusion}

The recent detection of QPOs in SGR (most likely associated with seismic vibrations triggered by magnetic crust quakes) 
and future observations with gravitational wave detectors, offer new possibilities to probe the interior of 
neutron stars and to test the theories of dense matter. 
Nevertheless the reliable identification of the various oscillation modes calls for a consistent theoretical description of 
the star. As a first step towards this goal, we have constructed fully self-consistent relativistic two-fluid models of neutron star cores, 
composed of superfluid neutrons and a conglomerate of protons, electrons and possibly muons. The mutual entrainment effects between the two fluids, 
resulting from the strong nucleon-nucleon interactions, are properly taken into account. 
We have determined the expression of the Lagrangian density in the variational framework developed by Brandon Carter and co-workers.
We have also shown how to make the correspondence with non-relativistic models by applying the 4D covariant 
formulation of Newtonian hydrodynamics of~\cite{CCI-04}. We have determined all the coefficients of these models consistently with the 
\emph{same} microscopic approach. In the perspective 
of describing not only the liquid core of neutron stars but also the crust layers, we have employed the nuclear 
energy density functional theory which has been already successfully applied to study both isolated nuclei and 
infinite nuclear matter. As an example, we have calculated the composition, the equation of state 
and the entrainment matrix of neutron star core for three different nuclear models: the popular SLy4 model for which the equation of state 
of both the crust and the liquid core has been already tabulated~\citep{haensel-04} and the more recent LNS and NRAPR models which have been entirely 
constructed from recent realistic many body calculations. For comparison, we have also considered relativistic $\sigma-\omega$ mean field models that have been 
first applied by \citet{comer-03}. We have improved their models by refitting the parameters in order to obtain a better agreement 
with nuclear data, but still we could not reach the same level of accuracy as the non-relativistic models mentioned above. This would require the introduction of additional meson fields, especially the rho meson. Besides self-meson couplings should be taken into account. 
Numerical calculations with both effective energy density functionals and relativistic mean field models have shown that the dynamical effective nucleon masses arising from entrainment effects 
are smaller than the ordinary mass in the Newtonian case. Relativistic effects increase effective masses, since all forms of internal energy contribute to the mass. A rather unexpected consequence which has not been usually discussed in the literature, is that relativistic effective masses can be even larger than the bare mass in the liquid core of neutron stars.

\section*{Acknowledgements}

This work was financially supported by FNRS (Belgium). The author is grateful to 
Brandon Carter for valuable comments during the completion of this work. 

\bibliography{paper}
\label{lastpage}
\end{document}